\documentclass[sigconf, balance=false]{acmart}
\renewcommand\footnotetextcopyrightpermission[1]{}
\usepackage{array,graphicx}
\usepackage{wasysym}
\usepackage[compact]{titlesec}

\usepackage{url}
\usepackage{indentfirst}
\usepackage{listings}
\usepackage{float}
\usepackage{subfig}
\usepackage{multirow}
\usepackage{soul}
\usepackage[notes=true,done=false]{dtrt}

\newcommand{\para}[1]{\vspace{0.75ex}\noindent{\bf \em #1}\hspace*{.3em}}

\title{Differential Degradation Vulnerabilities in \\ Censorship Circumvention Systems}

\author{Zhen Sun}
\affiliation{%
  \institution{Cornell Tech}
  \country{}}
\email{zs352@cornell.edu}

\author{Vitaly Shmatikov}
\affiliation{%
  \institution{Cornell Tech}
  \country{}}
\email{shmat@cs.cornell.edu}

\begin{document}

\begin{abstract}

Several recently proposed censorship circumvention systems use encrypted network channels of popular applications to hide their communications.  For example, a Tor pluggable transport called Snowflake uses the WebRTC data channel, while a system called Protozoa substitutes content in a WebRTC video-call application.  By using the same channel as the cover application and (in the case of Protozoa) matching its observable traffic characteristics, these systems aim to resist powerful network-based censors capable of large-scale traffic analysis.  Protozoa, in particular, achieves a strong indistinguishability property known as behavioral independence.

We demonstrate that this class of systems is generically vulnerable to a new type of active attacks we call ``differential degradation.'' These attacks do not require multi-flow measurements or traffic classification and are thus available to all real-world censors.   They exploit the discrepancies between the respective network requirements of the circumvention system and its cover application.  We show how a censor can use the minimal application-level information exposed by WebRTC to create network conditions that cause the circumvention system to suffer a much bigger degradation in performance than
the cover application.  Even when the attack causes no observable differences in network traffic and behavioral independence still holds, the censor can block circumvention at a low cost, without resorting to traffic analysis, and with minimal collateral damage to non-circumvention users. 

We present effective differential degradation attacks against Snowflake and Protozoa.  We explain the root cause of these vulnerabilities, analyze the tradeoffs faced by the designers of circumvention systems, and propose a modified version of Protozoa that resists differential degradation attacks.

\end{abstract}


\maketitle
\thispagestyle{plain}
\pagestyle{plain}

\section{Introduction}
\label{sec:intro}

Many governments around the world require local Internet providers to censor users' access to foreign news websites, online forums, and other sources of information.  In addition to blocking IP addresses of censored sites, network-based censors use deep packet inspection, protocol identification, and traffic analysis to detect and block attempts to evade censorship~\cite{censorship-survey}.

Research on censorship-resistant communications produced several systems that use the same network channels as ``cover'' application(s) which are permitted and popular in censorship regions.  Some even shape their traffic so that it looks, to a network observer, like the traffic of a cover application.  The key assumption behind this design is that censors want to avoid collateral damage and are thus not willing to block popular applications.

Early systems of this type attempted to imitate applications like Skype\cite{Moghaddam2012a}. Perfect imitation of complex interactive applications is not feasible, while imperfect imitation is easy to detect via tell-tale discrepancies in network traffic and responses to active probes~\cite{houmansadr2013parrot}.

\emph{Tunneled circumvention systems} avoid the flaws of imperfect imitation by actually executing the cover application and sending circumvention traffic through its network channel (e.g., substituting communication contents at the application or network level).  This ensures that the system is immune to imitation-detection attacks because it uses exactly the same protocol and implementation as the cover application.  Examples include FreeWave~\cite{houmansadr2013want}, CovertCast~\cite{mcpherson2016covertcast}, Protozoa~\cite{barradas2020poking}, Balboa~\cite{rosen2021balboa}, Telepath~\cite{sun2023telepath}, and Tor pluggable transports such as Snowflake~\cite{fifield2017threat}.

Even tunneled circumvention systems are vulnerable to detection if their network traffic is statistically distinguishable from the typical traffic of the cover application~\cite{barradas2018effective}.  For example, FreeWave~\cite{houmansadr2013want} is distinguishable
because it uses a voice channel but does not transmit actual speech~\cite{geddes2013cover}.

To evade detection via traffic analysis,
a circumvention system must be \emph{behaviorally independent}~\cite{raven-popets2022}: its network behavior must be indistinguishable from the cover application.  Protozoa~\cite{barradas2020poking} is named in~\cite{raven-popets2022} as an example of a behaviorally independent system.

The choice of a cover application is constrained
by several requirements~\cite{iv2022security,khattak2016sok}.  It must provide (1) an end-to-end encrypted channel to servers or peers outside the censorship region.  This channel must have (2) low latency and high bandwidth, to support interactive functionality such as Web browsing.  Critically, (3) the cover application must be popular among users in the censorship region.  The fundamental assumption is that a censor's attempts to block circumvention would cause collateral damage to the non-circumvention users of the cover application.  Presumably, these users greatly outnumber those who engage in circumvention, and it is assumed that censors are not willing to annoy them.   If a circumvention system uses an unpopular cover application, censors can simply block all network flows that appear to belong to the application in question, without needing to distinguish if they carry genuine application traffic or tunneled circumvention traffic.

Video applications emerged in the research literature as a common choice that satisfies all three requirements.  They provide high-capacity encrypted channels and are often available in open source, enabling modification and integration with circumvention systems.  Modern interactive video applications typically use WebRTC~\cite{webrtc,rtcweb} as the underlying network protocol, since WebRTC was specifically designed for real-time communications.

\para{Our contributions.}
We demonstrate that systems that are ``behaviorally independent'' in the sense of~\cite{raven-popets2022} can still be blocked using simple techniques that are readily available to real-world censors and require no traffic analysis or classification.

Our first observation is that the network requirements of circumvention systems do not match those of cover applications.  One may require an asynchronous channel, the other synchronous; one may need bidirectionality, while the other sends traffic only in one direction.  Reliability and quality-of-service requirements don't match, either.  These discrepancies are usually caused by differences in functionality, such as video streaming vs.\ interactive chat vs.\ Web browsing.  For example, interactive, WebRTC-based applications like Discord and Google Meet need a reliable asynchronous data channel (for peer-to-peer chat); Snowflake \cite{Bocovich2024a}, a Tor pluggable transport which uses WebRTC for cover, needs a reliable synchronous channel (for client-server communications).

Our second observation is that encrypted channels used by tunneled circumvention systems do not hide all application-specific information.  The revealed metadata may not distinguish circumvention and cover traffic (thus behavioral independence holds), but the censor can identify the cover application and use their knowledge of the application's semantics to create network conditions that cause predictable changes in the application-level behavior.

Our third, most important observation is that \textbf{detection is not necessary for blocking}.  Behavioral independence is an indistinguishability property, defined over the censor's view of network traffic. It aims to prevent detection (not blocking!) by hiding the differences between the traffic of the circumvention system and the cover application.  If the censor-imposed network conditions damage the former more than the latter, the censor can block circumvention\textemdash even if behavioral independence holds and the effects of the attack are not visible at the network level.

We present a new class of active attacks we call \textbf{differential degradation}.  They enable network-based adversaries to block tunneled circumvention systems with minimal collateral damage to cover applications.  These attacks are straightforward to deploy because the censor only needs to inspect and selectively drop individual packets, control coarse channel characteristics such as bandwidth and latency, and, at most, perform basic single-flow measurements such as computing the average packet size over some period.  The censor does not need to collect or label multi-flow data, nor train any single- or multi-flow classifiers.

The censor first identifies the application (in WebRTC, this information is available from unencrypted headers), then uses unencrypted application-level features revealed by the protocol (e.g., video-frame markers) to selectively block some channels, or else impose network conditions that elicit different application-level adaptation in the circumvention system and the cover application.  By causing full disruption or significant performance degradation in the circumvention system\textemdash while the cover application is unaffected or experiences a much smaller degradation\textemdash these attacks effectively block circumvention with minimal collateral damage and \emph{without distinguishing circumvention and cover traffic}.  This breaks the key assumption of tunneled circumvention systems.

Unlike ``Dead Parrot''~\cite{houmansadr2013parrot} and ``Cover Your ACKS''~\cite{geddes2013cover} attacks, differential degradation attacks do not exploit imperfect imitation.  They work even against behaviorally independent systems like Protozoa that produce statistically indistinguishable traffic.

We then discuss the tradeoffs faced by the designers of tunneled circumvention systems and explain how their choices are constrained by the available cover applications.  Popular interactive, bidirectional video applications are all based on WebRTC.  Circumvention systems (such as Protozoa) that use WebRTC for cover are inherently vulnerable to differential degradation attacks.  

The alternative is TLS-based cover applications.
We implement and evaluate Ciliate, a version of Protozoa that runs over TLS-encrypted video-streaming and resists differential degradation.  TLS-based video applications, however, only provide unidirectional channels.  Implementing interactive functionalities such as Web browsing thus requires multiple channels, potentially exposing Ciliate (and similar systems such as Balboa~\cite{rosen2021balboa}) to detection due to anomalous network flows.  This detection requires multi-flow traffic analysis and accurate models of normal network flows.  By contrast, differential degradation is a simple single-flow attack, available even to basic network-based censors.

Given popular high-bandwidth cover applications that are available today, this tradeoff is fundamental: WebRTC-based tunneled circumvention systems are vulnerable to differential degradation, TLS-based systems potentially introduce anomalous network flows.

\section{Background}

We first explain the basic principles of tunneled circumvention, define behavioral independence, and survey available cover applications.  We then summarize WebRTC, a real-time media transmission protocol used by many cover applications.  Finally, we describe Snowflake and Protozoa, two representative systems that we focus on in the rest of this paper.

\subsection{Tunneled circumvention systems}

\begin{figure}[h]
\centering
\includegraphics[width=\linewidth]{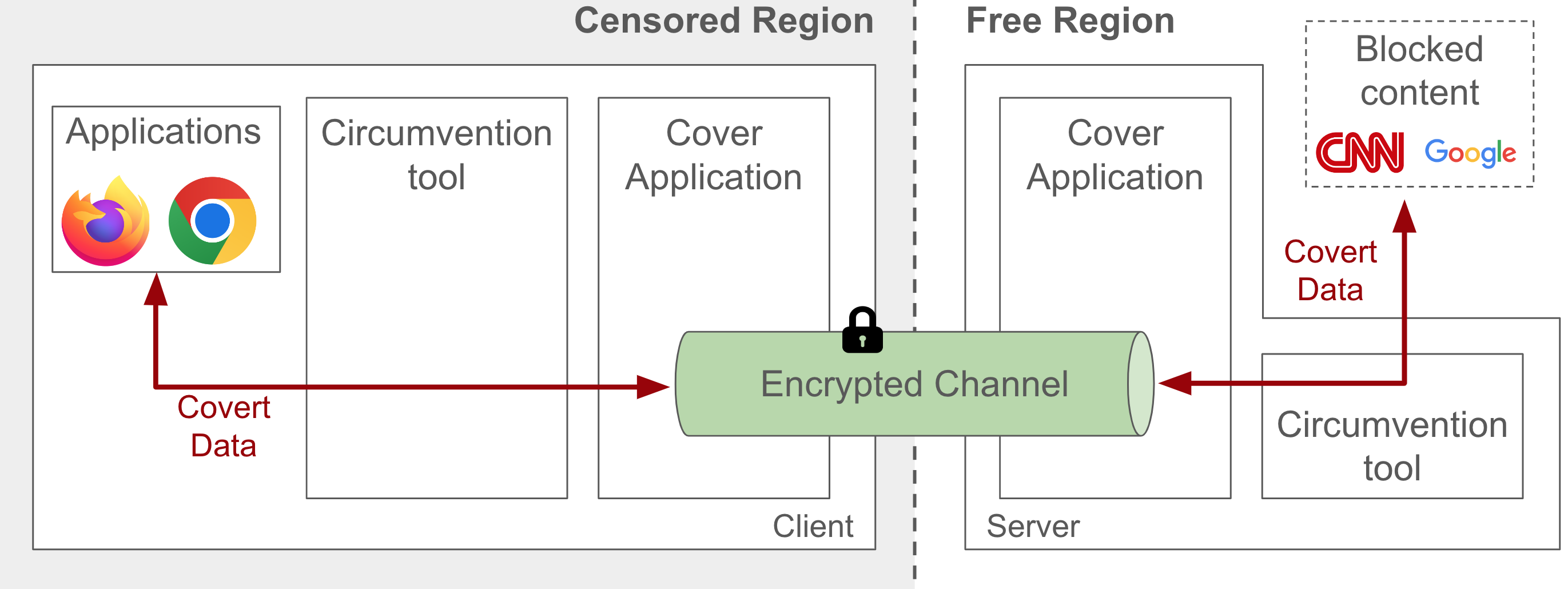}
\centering
\caption{A tunneled circumvention system.}
\label{fig:tunneled}
\end{figure}

The primary goal of tunneled circumvention systems is to make their communications look similar to some ``cover'' application.  Given the threat model of network-based Internet censorship (see Section~\ref{sec:threat}), censors may be capable of deploying machine learning-based traffic analysis to detect differences between the respective traffic patterns of the circumvention system and the cover application, even if both use the same protocol~\cite{barradas2018effective}.  Systems that aim to resist traffic analysis need to match all observable features of the cover application's network traffic, including distributions of packet sizes, packet counts in both directions, inter-packet intervals, etc.  Because it is difficult to faithfully imitate the behavior of a complex network stack~\cite{houmansadr2013parrot}, traffic analysis-resistant circumvention systems actually execute the cover application and use its network channel(s) to tunnel circumvention traffic (see Figure~\ref{fig:tunneled}).  

A common type of tunneled circumvention systems~\cite{mcpherson2016covertcast, barradas2020poking, rosen2021balboa, sun2023telepath} relies on \emph{content substitution}.  It replaces the data sent by the cover application either at the application level (e.g., by encoding it into the video content~\cite{mcpherson2016covertcast}) or at the network level, before it is transmitted by the application's network manager~\cite{barradas2020poking, rosen2021balboa}.

\subsection{Behavioral independence}

Behavioral independence, defined in \cite{raven-popets2022}, requires that the observable behavior of the circumvention channel be indistinguishable from the genuine cover-application behavior regardless of the channel input.  The scope of ``behavior'' is not fully specified in~\cite{raven-popets2022}, but Protozoa~\cite{barradas2020poking} is given as an example of behavioral independence.  Network censors can only observe the system's network behavior, thus behavioral independence is a \emph{channel} property (as opposed to an application property). Network-level behavioral independence is also achieved by Balboa~\cite{rosen2021balboa} and Telepath~\cite{sun2023telepath}.

In general, resistance to statistical traffic analysis is not sufficient to achieve even network-level behavioral independence.  Depending on the cover application, content substitution may produce traffic traces that are statistically indistinguishable from the traffic of the cover application, yet violate application-specific causal dependencies.  These traces are easily detectable because they are impossible in the cover application~\cite{sun2023telepath}.  In this paper, we focus on a different class of attacks that do not rely on detection and affect even systems that are truly behaviorally independent at the network level.

Behavioral realism, as defined in \cite{raven-popets2022}, requires that human-level behavior of the circumvention system be indistinguishable from the genuine behavior of the cover application's users.  Neither behavioral independence, nor behavioral realism
considers \emph{unobservable} application-level behavior.

\subsection{Cover applications}
\label{sec:cover}

Examples of cover applications that provide high-capacity,
end-to-end encrypted channels and are popular in censorship regions can be found in~\cite{iv2022security,khattak2016sok}.
\emph{Streaming media} applications are natural candidates, but different types of streaming applications use
network channels with substantially different properties.

Interactive streaming applications typically use \textbf{real-time, bidirectional, UDP-based, unreliable channels}.   FreeWave~\cite{houmansadr2013want} implements a reliable TCP channel over Skype audio but produces detectable traffic~\cite{geddes2013cover}. SkypeLine~\cite{kohls2016skypeline} handles channel mismatches but provides only relatively low bandwidth.
DeltaShaper~\cite{barradas2017deltashaper} produces traffic patterns distinct from Skype video and is thus detectable. Protozoa~\cite{barradas2020poking} uses WebRTC video and resists traffic analysis. 

WebRTC (see Section~\ref{sec:webrtc}) is a common choice because it is easy to modify and integrate. 
WebRTC is used by popular applications such as Discord, Facebook Messenger, and Google Chat/Meet, thus censors may be reluctant to block WebRTC due to collateral damage.

Other streaming applications use \textbf{asynchronous, unidirectional, TCP-based, reliable} channels. Facet~\cite{li2014facet} streams videos through the Skype video-chat channel.  CovertCast~\cite{mcpherson2016covertcast} uses RTMP to stream video-encoded Web content through YouTube. Balboa~\cite{rosen2021balboa} provides a unidirectional channel over HTTP or audio streaming.

\emph{Online games} are another category of cover applications.  Examples of game-based circumvention systems include Castle~\cite{hahn2016games} (a real-time strategy game), Rook \cite{vines2015rook} (a first-person shooter game), and Telepath~\cite{sun2023telepath} (based on Minecraft).  Game channels have much lower capacity than streaming media.  Furthermore, games have complex application semantics, including state machines that are partially observable even in encrypted network traffic~\cite{sun2023telepath}.

\subsection{WebRTC}
\label{sec:webrtc}

Web Real-Time Communication (WebRTC) is a peer-to-peer framework supported by Web browsers and mobile applications~\cite{webrtc,rtcweb}. WebRTC enables peer-to-peer (P2P) audio, video, and data communication without plugins or other user-installed software. 

\begin{figure}[h]
\centering
\includegraphics[width=\linewidth]{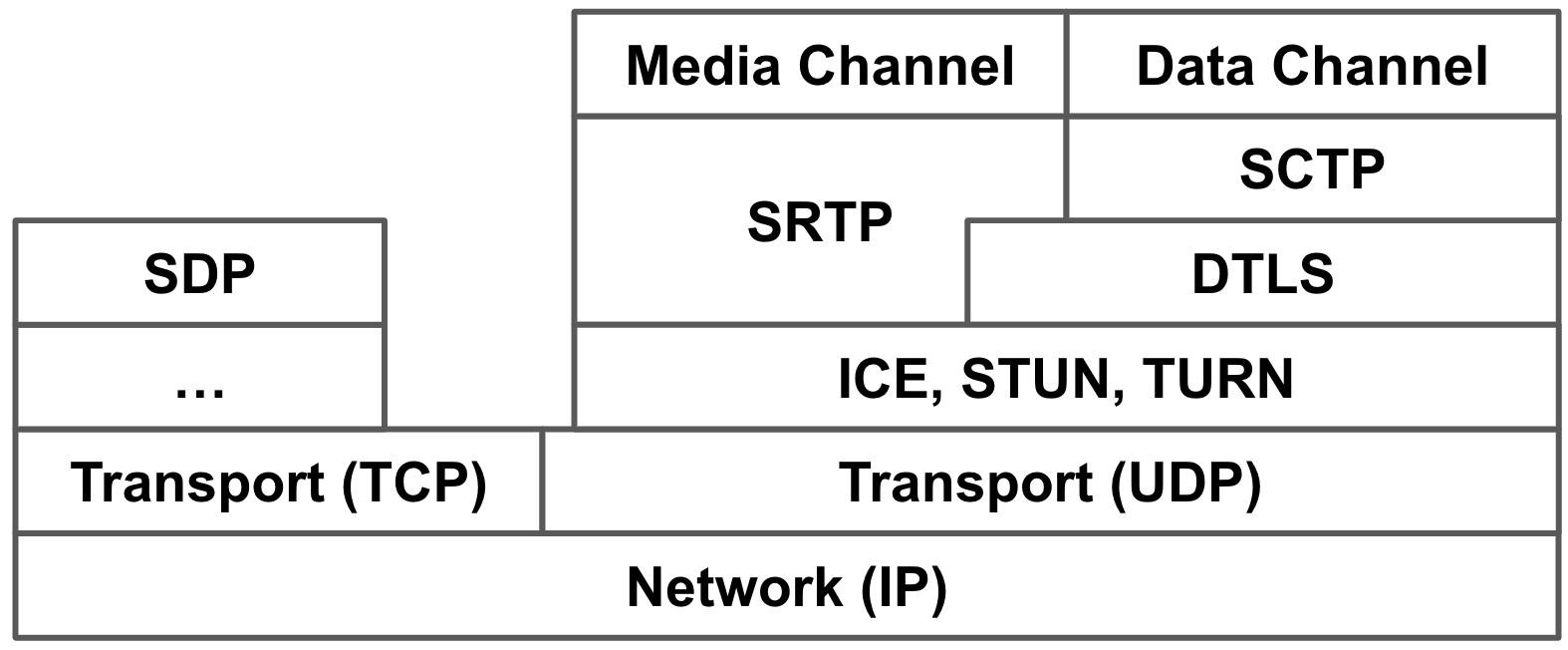}
\centering
\caption{WebRTC protocol stack.}
\label{fig:webrtc-protocol}
\end{figure}

Figure \ref{fig:webrtc-protocol} shows the WebRTC protocol stack.  Peers first perform the signaling process (not provided by WebRTC and usually implemented by the application developer) to exchange bootstrapping information using Session Description Protocol (SDP). WebRTC then uses Interactive Connectivity Establishment (ICE) to set up P2P connections. ICE uses STUN~\cite{stun} and TURN~\cite{turn} for NAT traversal.  Over the established P2P connection, WebRTC provides two types of encrypted channels: media channels and data channels. 

\para{Media channel.} 
Media channel enables webpages and applications to access media streams (e.g., feeds from a camera or microphone) and lets peers exchange media data in real time using Secure Real-time Transport Protocol (SRTP)~\cite{srtp}. 

The key establishment protocol is DTLS-SRTP~\cite{dtls-srtp}.  Once the peers set up an ICE session, they perform a DTLS handshake with the “use\_srtp” extension. After the DTLS session is established, the peers use their DTLS keys to derive SRTP keys.  Media data are encrypted using SRTP only and never sent in DTLS record-layer “application\_data” packets, i.e., the media channel uses DTLS only for key exchange and not for actual communication.

\para{Data channel.}
Data channel is an encrypted, reliable communication channel that lets webpages and applications directly send non-media data to clients.  Data are sent using SCTP~\cite{dtls-sctp} encapsulated in the DTLS session established via the media-channel DTLS-SRTP key exchange. The DTLS layer provides encryption, while the SCTP layer provides reliability.  All data-channel communications are sent in DTLS record-layer “application\_data” packets.

\subsection{Snowflake}
\label{sec:snowflake-overview}

\begin{figure}[h]
\centering
\includegraphics[width=\linewidth]{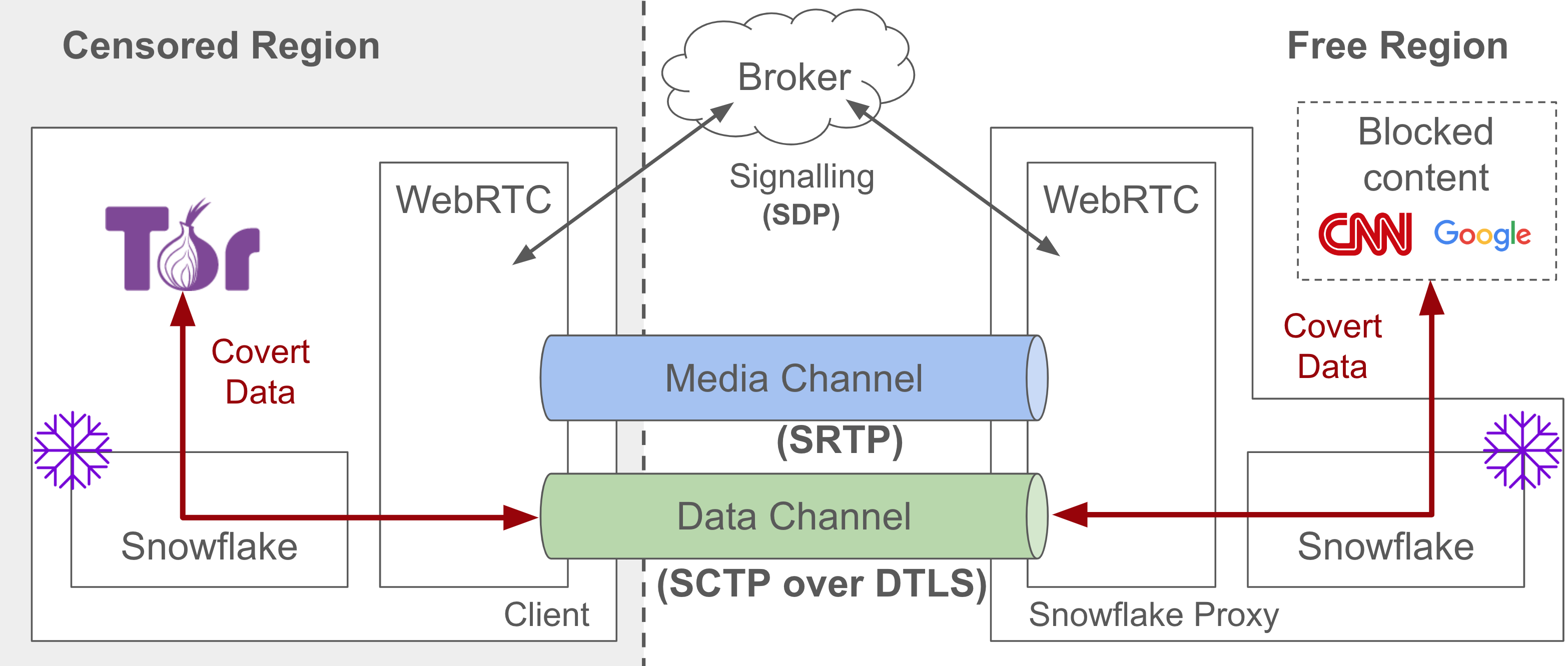}
\centering
\caption{Overview of Snowflake.}
\label{fig:snowflake}
\end{figure}

Snowflake \cite{Bocovich2024a} is a Tor~\cite{dingledine2004tor} pluggable transport~\cite{tor-pluggable} that uses the WebRTC data channel to evade Tor blocking\textemdash see Figure \ref{fig:snowflake}. 

Users in the censorship region run Snowflake clients that connect to a volunteer-operated, in-browser Snowflake proxy outside the censorship region using a P2P connection over the WebRTC data channel.  Tor data are tunneled through this channel.  Snowflake relies on a broker running on a third-party Web service to match clients with proxies and to exchange bootstrapping information.

\subsection{Protozoa}
\label{sec:protozoa-overview}

Protozoa is a tunneled circumvention system that uses the WebRTC media channel.  Unlike Snowflake, Protozoa aims to resist statistical traffic analysis, including machine learning-based classifiers, and achieves behavioral independence~\cite{raven-popets2022}.

\begin{figure}[h]
\centering
\includegraphics[width=\linewidth]{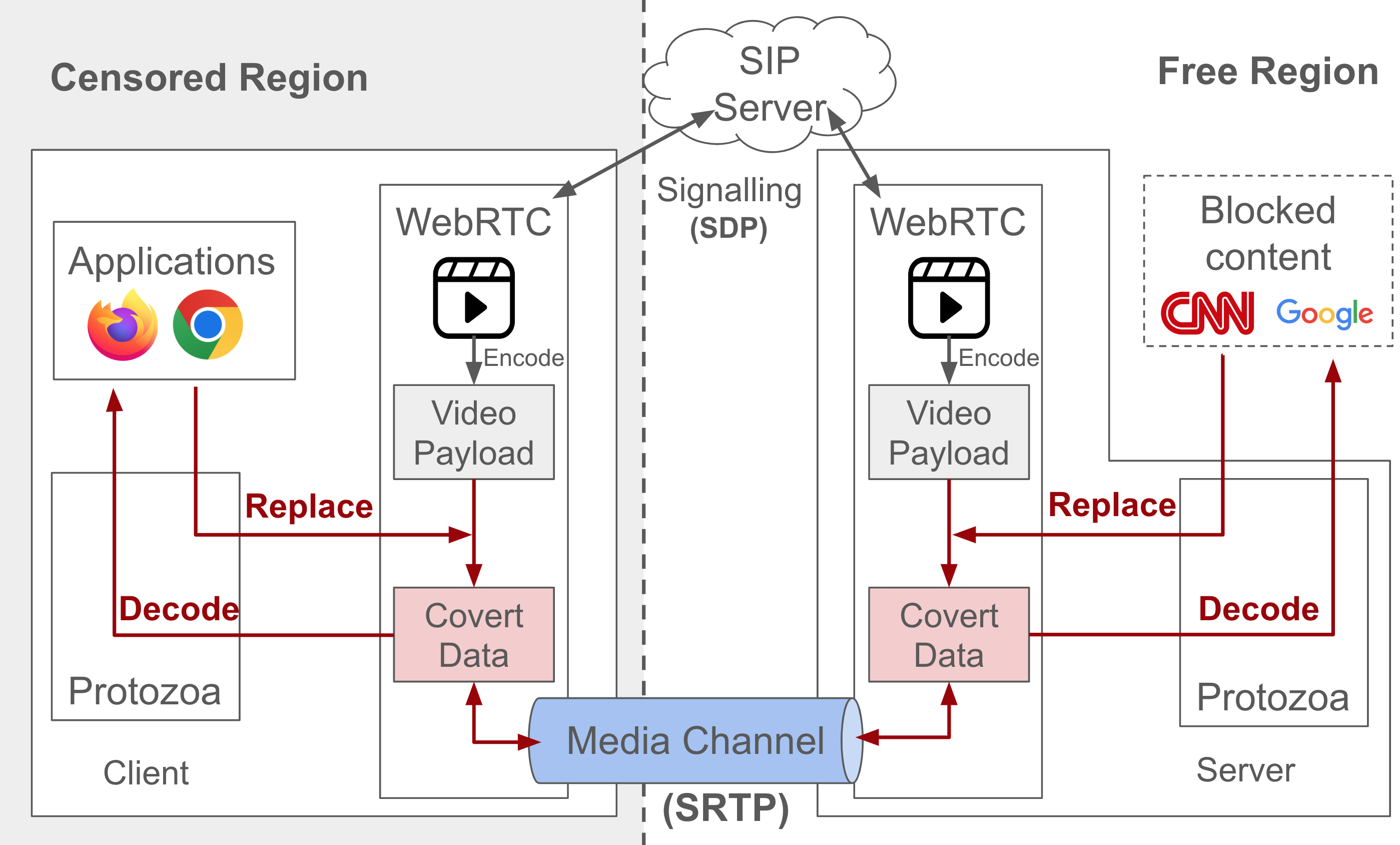}
\centering
\caption{Overview of Protozoa.}
\label{fig:protozoa}
\end{figure}

Figure \ref{fig:protozoa} shows a schematic overview of Protozoa.  The key idea is to run the entire video-streaming application and substitute the encoded video payload with circumvention data.  The advantage of this design is that network traffic generated by Protozoa preserves the statistical characteristics, such as packet sizes and inter-packet intervals, of a real video stream. Protozoa requires the client and the proxy to agree on a common rendezvous point (a SIP server, e.g., a video-chat room) to exchange bootstrapping information.

\section{Censorship Models}
\label{sec:threat}

In this section, we discuss censorship models considered in the circumvention literature and compare them to real-world censors.  We then show that our attack assumes a realistic censorship model.  Our classification rubrics follow~\cite{houmansadr2013parrot}.

\subsection{Censors' capabilities}

A \textbf{passive} adversary operates at the ISP level and has full access to all network traffic transiting the ISP's routers (e.g., using traffic-interception equipment deployed within the ISP).  We assume that this adversary can perform deep packet inspection, i.e., read \emph{unencrypted} contents of all network packets beyond IP headers.  We also assume that encryption is secure and hides all information about encrypted data except its size.  

Tunneled circumvention systems usually assume that encryption hides all application-level information.  Recent research~\cite{sun2023telepath} and the attacks presented in this paper demonstrate, however, that application states and behavior can be inferred from the \textbf{metadata}, including unencrypted parts of packet headers.  Inferring application-level information from this metadata is straightforward and can be performed even by basic censors at line speeds.

Adversaries who have access to decryption keys (by compromising the endpoints or via key escrow) and/or are capable of breaking key establishment protocols (e.g., via man-in-the-middle attacks on TLS connections using compromised public-key certificates) are outside our threat model.

An \textbf{active} or \textbf{traffic manipulation} adversary can modify traffic flows.  This adversary can drop selected packets, throttle the bandwidth, or increase latency by storing and re-transmitting packets.

A \textbf{proactive} adversary can initiate new connections to suspected hosts and send active probes to them, unlike traffic-manipulation adversary that only perturbs existing connections.

\subsection{Censors' knowledge}

A \textbf{constant-packet} adversary inspects individual packets belonging to a particular \emph{flow} (a pair of communicating hosts) but does not perform any measurements that involve multiple packets. 

A \textbf{single-flow} adversary is slightly more sophisticated.  In addition to inspecting individual packets, he can perform coarse longitudinal measurements over a single flow (e.g., compute average packet size, measure if it changes over time, etc.).  These measurements are limited to simple statistics and do not involve applying machine-learning classifiers.

A \textbf{multi-flow traffic analysis} adversary is the standard threat assumed by the latest tunneled circumvention systems~\cite{barradas2020poking, rosen2021balboa, sun2023telepath}.  This adversary can collect and label large volumes of circumvention and non-circumvention traffic generated by a variety of applications, then use machine learning to train traffic classifiers that take advantage of all available features (packet counts, sizes, inter-packet intervals, etc.) to try and distinguish circumvention and cover traffic without a prohibitive number of false positives.

\begin{table*}[t]
\small
\newcommand{\dummy}{\rule[-2em]{0pt}{2em}}
\begin{tabular}{|ll|ccccccc|ccc|c|}
\hline
\multicolumn{2}{|l|}{\multirow{3}{*}{}}                                                                                                                                                                                                & \multicolumn{7}{c|}{Research papers}                                                                                                                                                                                                                                                                                                                                                                                                                                                                                             & \multicolumn{3}{c|}{Real censors}                                                                                                                                                                                                                         & \multirow{3}{*}{\begin{tabular}[c]{@{}c@{}}\textbf{Differential} \\ \textbf{degradation} \\ (this paper) \end{tabular}} \\ \cline{3-12}
\multicolumn{2}{|l|}{}                                                                                                                                                                                                                 & \multicolumn{4}{c|}{\begin{tabular}[c]{@{}c@{}}Evaluation of \\ existing systems\end{tabular}}                                                                                                                                                                                                                                                                                                   & \multicolumn{3}{c|}{\begin{tabular}[c]{@{}c@{}}Attacks on \\ existing systems\end{tabular}}                                   & \multicolumn{1}{c|}{\begin{tabular}[c]{@{}c@{}}Blocking \\ Tor\end{tabular}} & \multicolumn{1}{c|}{\begin{tabular}[c]{@{}c@{}}Blocking \\ Shadowsocks\end{tabular}} & \begin{tabular}[c]{@{}c@{}}Blocking \\ VMess, \\ Obfs4, \\ Shadowsocks\end{tabular} &                                                                                                                         \\ \cline{3-12}
\multicolumn{2}{|l|}{}                                                                                                                                                                                                                 & \multicolumn{1}{c|}{\begin{tabular}[c]{@{}c@{}}Protozoa\\ \cite{barradas2020poking}\end{tabular}} & \multicolumn{1}{c|}{\begin{tabular}[c]{@{}c@{}}Balboa\\ \cite{rosen2021balboa}\end{tabular}} & \multicolumn{1}{c|}{\begin{tabular}[c]{@{}c@{}}Telepath\\ \cite{sun2023telepath}\end{tabular}} & \multicolumn{1}{c|}{\begin{tabular}[c]{@{}c@{}}Raven\\ \cite{raven-popets2022}\end{tabular}} & \multicolumn{1}{c|}{\cite{barradas2018effective}} & \multicolumn{1}{c|}{\cite{geddes2013cover}} & \cite{houmansadr2013parrot} & \multicolumn{1}{c|}{\cite{Tschantz2016a,Winter2012a,Ensafi2015b}}            & \multicolumn{1}{c|}{\cite{Alice2020a}}                                               & \cite{Wu2023a}                                                                      &                                                                                                                         \\ \hline
\multicolumn{1}{|l|}{\multirow{3}{*}{\rotatebox[origin=c]{90}{\begin{tabular}[c]{@{}l@{}}Multi-flow \\ measurement\end{tabular}}}}                      & \begin{tabular}[c]{@{}l@{}}Data \\ collection\\ and \\ labeling\end{tabular} & \multicolumn{1}{c|}{\CIRCLE}                                                                      & \multicolumn{1}{c|}{\CIRCLE}                                                                 & \multicolumn{1}{c|}{\CIRCLE}                                                                   & \multicolumn{1}{c|}{\CIRCLE}                                                                 & \multicolumn{1}{c|}{\CIRCLE}                      & \multicolumn{1}{c|}{\LEFTcircle}            & \Circle                     & \multicolumn{1}{c|}{\Circle}                                                 & \multicolumn{1}{c|}{\Circle}                                                         & \Circle                                                                             & \Circle                                                                                                                 \\ \cline{2-13} 
\multicolumn{1}{|l|}{}                                                                                                                                  & \begin{tabular}[c]{@{}l@{}}+ Statistical\\ test\end{tabular}                 & \multicolumn{1}{c|}{\Circle}                                                                      & \multicolumn{1}{c|}{\Circle}                                                                 & \multicolumn{1}{c|}{\Circle}                                                                   & \multicolumn{1}{c|}{\Circle}                                                                 & \multicolumn{1}{c|}{\LEFTcircle}                  & \multicolumn{1}{c|}{\LEFTcircle}            & \Circle                     & \multicolumn{1}{c|}{\Circle}                                                 & \multicolumn{1}{c|}{\Circle}                                                         & \Circle                                                                             & \Circle                                                                                                                 \\ \cline{2-13} 
\multicolumn{1}{|l|}{}                                                                                                                                  & \begin{tabular}[c]{@{}l@{}}+ ML\\ classifier\end{tabular}                    & \multicolumn{1}{c|}{\CIRCLE}                                                                      & \multicolumn{1}{c|}{\CIRCLE}                                                                 & \multicolumn{1}{c|}{\CIRCLE}                                                                   & \multicolumn{1}{c|}{\CIRCLE}                                                                 & \multicolumn{1}{c|}{\LEFTcircle}                  & \multicolumn{1}{c|}{\Circle}                & \Circle                     & \multicolumn{1}{c|}{\Circle}                                                 & \multicolumn{1}{c|}{\Circle}                                                         & \Circle                                                                             & \Circle                                                                                                                 \\ \hline
\multicolumn{1}{|l|}{\multirow{3}{*}{\rotatebox[origin=c]{90}{\begin{tabular}[c]{@{}l@{}}Single-flow/ \\ constant-packet \\ measurement\end{tabular}}}} & \begin{tabular}[c]{@{}l@{}}Plaintext/\\ metadata\\ filtering\end{tabular}    & \multicolumn{1}{c|}{\frownie}                                                                     & \multicolumn{1}{c|}{\textbf{?}}                                                              & \multicolumn{1}{c|}{\LEFTcircle}                                                               & \multicolumn{1}{c|}{\textbf{?}}                                                              & \multicolumn{1}{c|}{\Circle}                      & \multicolumn{1}{c|}{\Circle}                & \LEFTcircle                 & \multicolumn{1}{c|}{\CIRCLE}                                                 & \multicolumn{1}{c|}{\Circle}                                                         & \LEFTcircle                                                                         & \CIRCLE                                                                                                                 \\ \cline{2-13} 
\multicolumn{1}{|l|}{}                                                                                                                                  & \begin{tabular}[c]{@{}l@{}}Threshold\\ classifer\end{tabular}                & \multicolumn{1}{c|}{\textbf{?}}                                                                   & \multicolumn{1}{c|}{\textbf{?}}                                                              & \multicolumn{1}{c|}{\textbf{?}}                                                                & \multicolumn{1}{c|}{\textbf{?}}                                                              & \multicolumn{1}{c|}{\Circle}                      & \multicolumn{1}{c|}{\Circle}                & \Circle                     & \multicolumn{1}{c|}{\Circle}                                                 & \multicolumn{1}{c|}{\LEFTcircle}                                                     & \LEFTcircle                                                                         & \Circle                                                                                                                 \\ \cline{2-13} 
\multicolumn{1}{|l|}{}                                                                                                                                  & \begin{tabular}[c]{@{}l@{}}Entropy \\ test\end{tabular}                      & \multicolumn{1}{c|}{\textbf{?}}                                                                   & \multicolumn{1}{c|}{\textbf{?}}                                                              & \multicolumn{1}{c|}{\textbf{?}}                                                                & \multicolumn{1}{c|}{\textbf{?}}                                                              & \multicolumn{1}{c|}{\Circle}                      & \multicolumn{1}{c|}{\Circle}                & \Circle                     & \multicolumn{1}{c|}{\Circle}                                                 & \multicolumn{1}{c|}{\LEFTcircle}                                                     & \CIRCLE                                                                             & \Circle                                                                                                                 \\ \hline
\multicolumn{2}{|l|}{Traffic manipulation}                                                                                                                                                                                             & \multicolumn{1}{c|}{\frownie}                                                                     & \multicolumn{1}{c|}{\textbf{?}}                                                              & \multicolumn{1}{c|}{\textbf{?}}                                                                & \multicolumn{1}{c|}{\textbf{?}}                                                              & \multicolumn{1}{c|}{\Circle}                      & \multicolumn{1}{c|}{\LEFTcircle}            & \LEFTcircle                 & \multicolumn{1}{c|}{\LEFTcircle}                                             & \multicolumn{1}{c|}{\Circle}                                                         & \Circle                                                                             & \CIRCLE                                                                                                                 \\ \hline
\multicolumn{2}{|l|}{Proactive attacks}                                                                                                                                                                                                & \multicolumn{1}{c|}{\textbf{?}}                                                                   & \multicolumn{1}{c|}{\textbf{?}}                                                              & \multicolumn{1}{c|}{\textbf{?}}                                                                & \multicolumn{1}{c|}{\textbf{?}}                                                              & \multicolumn{1}{c|}{\Circle}                      & \multicolumn{1}{c|}{\Circle}                & \LEFTcircle                 & \multicolumn{1}{c|}{\LEFTcircle}                                             & \multicolumn{1}{c|}{\LEFTcircle}                                                     & \LEFTcircle                                                                         & \Circle                                                                                                                 \\ \hline
\multicolumn{2}{|l|}{Requires detection?}                                                                                                                                                                                                  & \multicolumn{1}{c|}{\CIRCLE}                                                                      & \multicolumn{1}{c|}{\CIRCLE}                                                                 & \multicolumn{1}{c|}{\CIRCLE}                                                                   & \multicolumn{1}{c|}{\CIRCLE}                                                                 & \multicolumn{1}{c|}{\CIRCLE}                      & \multicolumn{1}{c|}{\LEFTcircle}            & \CIRCLE                     & \multicolumn{1}{c|}{\CIRCLE}                                                 & \multicolumn{1}{c|}{\CIRCLE}                                                         & \Circle                                                                             & \LEFTcircle                                                                                                             \\ \hline
\end{tabular}

\label{table:compare}

\caption{Censorship models of recently proposed circumvention systems, attacks against existing systems, and real censors. \CIRCLE \, denotes that the attack or the evaluation of a system assumes this adversary type. \LEFTcircle \, denotes that the attack partly assumes this adversary type. \Circle \, denotes that the attack does not assume this adversary type. \textbf{?} \, denotes that the system was not evaluated against this adversary type. \frownie \, denotes that the system is vulnerable to this adversary type.}
\label{table:censorship}
\end{table*}

\subsection{Real censors vs. research studies}

Table \ref{table:censorship} summarizes the main censorship techniques considered in recently published circumvention systems, attacks, and measurement studies of real-world censors.

There is a significant gap between real-world censorship techniques and hypothetical censors considered
in the circumvention literature (this gap was also observed in~\cite{Tschantz2016a}).   Real censors prefer cheap techniques such as single-flow or constant packet passive analysis, traffic manipulation, and proactive attacks. Research literature focuses on expensive, multi-flow traffic analysis.

State-of-the-art attacks on tunneled circumvention~\cite{barradas2018effective} assume a multi-flow traffic analysis adversary that collects network traces generated by cover applications and circumvention systems, labels the data, and then trains statistical tests and machine-learning based classifiers using hundreds of features extracted from the data to reliably identify the difference in traffic patterns such as packet sizes and timings.  Recently published tunneled circumvention systems such as Protozoa~\cite{barradas2020poking} and Balboa~\cite{rosen2021balboa} use similar multi-flow traffic analysis classifiers to evaluate their behavioral independence. Telepath~\cite{sun2023telepath} and Raven~\cite{raven-popets2022} assume adversaries that are aware of application semantics and human behavior, respectively, but both systems only consider multi-flow adversaries in their evaluation.

Although there has been some research on efficient multi-flow traffic analysis at large scale~\cite{barradas2021flowlens}, there is no concrete evidence that such attacks are feasible in the real world, nor have they been adopted by censors so far.  Measurement studies of how censors block systems like Tor, Shadowsocks, and VMess show that they prefer cheap, simple techniques~\cite{Tschantz2016a,Winter2012a,Ensafi2015b}. For example, to block Shadowsocks, the Great Firewall of China (GFW) first identifies suspected Shadowsocks servers using a single-flow threshold classifier and entropy test, then actively probes them~\cite{Alice2020a}.  For the recent blocking of fully encrypted channels, GFW uses a even cheaper technique: it blocks all traffic that fails a single-flow entropy test (with a few exemption rules).  This effectively blocks VMess, obfs4, and Shadowsocks with few false positives~\cite{Wu2023a}.  This attack is a pure blocking attack that does not require detection, and a real-world example of a channel mismatch.

Differential degradation attacks demonstrated in the rest of this paper can be deployed by constant-packet and single-flow adversaries using passive measurement and traffic manipulation, which are close to real-world censors and much easier than the multi-flow traffic analysis adversaries assumed by the recent literature.

\section{Differential Degradation Attacks}

Tunneled censorship circumvention systems use the same network channel as some cover application and\textemdash to avoid imitation\textemdash rely on the latter to provide the sizes and timings of observable network activities (e.g., they may substitute the contents of cover application's communications after they are sent to the network manager but before they are transmitted).  In normal operation, this helps circumvention systems produce traffic that is indistinguishable from the traffic produced by the cover application.

Even so, a circumvention system and a cover application that are indistinguishable \emph{at the network level} may not make the same assumptions about the network and thus may not react or adapt in the same way to changing network conditions \emph{at the application level}.  Differential degradation attacks exploit these discrepancies.  

\begin{figure}[h]
\centering
\includegraphics[width=\linewidth]{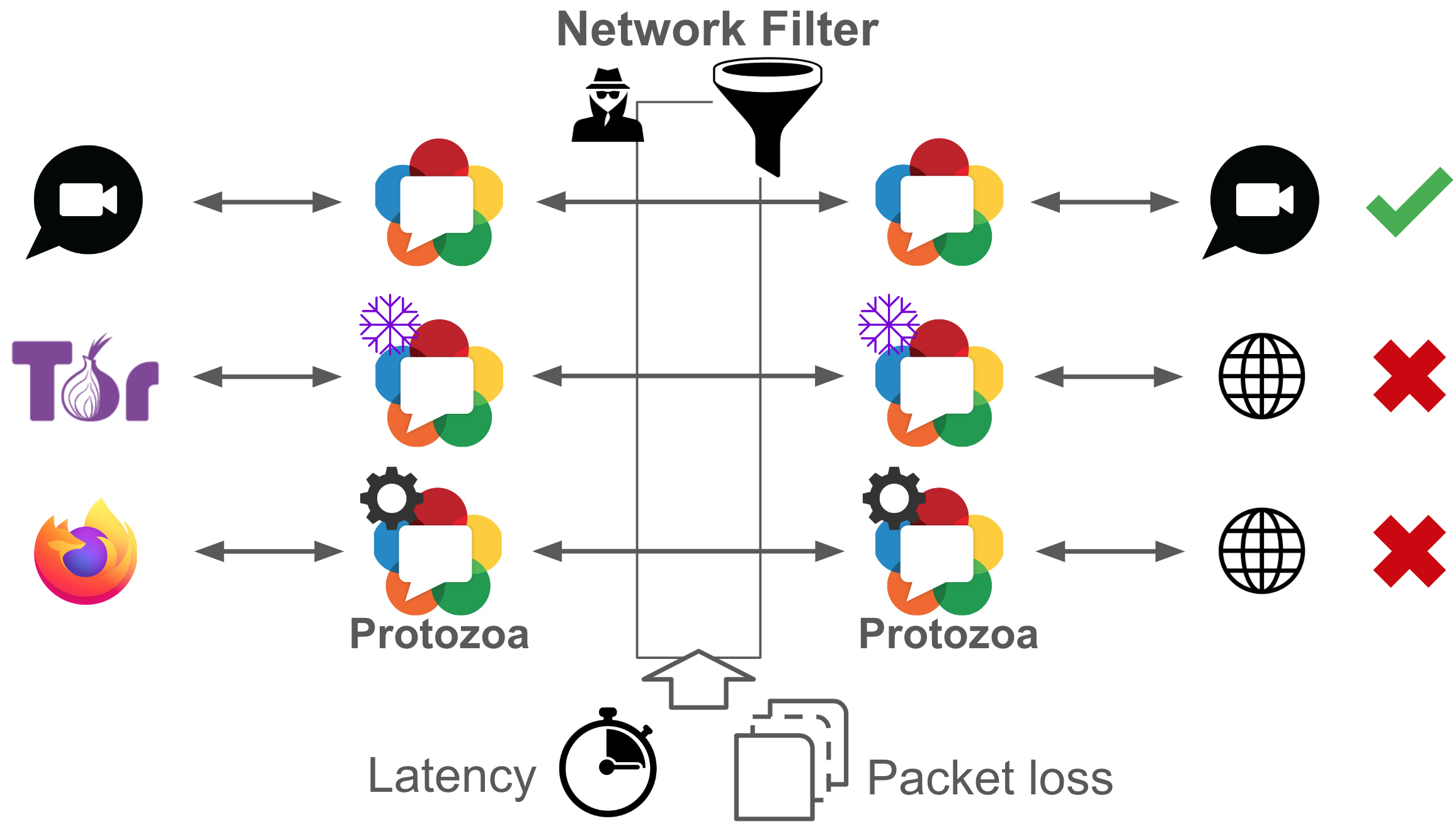}
\centering
\caption{Differential degradation attacks.}
\label{fig:attack-overview}
\end{figure}

Figure~\ref{fig:attack-overview} gives a high-level overview of differential degradation attacks.  They leverage three observations.  

First, protocols (notably, WebRTC) used by low-latency, high-bandwidth cover applications reveal some application-level information in unencrypted headers.  A network-based adversary can tell which application generated the traffic in question and also see some coarse content-related data, such as video-frame boundaries.  By itself, this information is not sufficient to distinguish circumvention and cover traffic and break behavioral independence.

Second, in many cases (see Section~\ref{sec:tradeoffs}), the functionality of circumvention systems (e.g., Web browsing) is fundamentally different from the functionality of cover applications (e.g., video calls).  Functionality determines which channels the respective systems use or, if they use the same channels, how they adapt when network conditions become worse.  This enables a network-based adversary to use his knowledge of the specific application (revealed by WebRTC application headers) to create network conditions that elicit different application-level behavior from the circumvention system and the cover application.

Third, \emph{an attack need not cause visible changes in traffic patterns to be effective}.  This distinguishes differential degradation from traffic analysis and other detection attacks, and explains why it is effective even against behaviorally independent systems.  The adversary's ultimate objective is to block circumvention without blocking cover applications.  Detection is useful insofar as it helps identify flows to block, but not necessary if the adversary can block without detecting\textemdash even he cannot directly observe the results of his actions.

To stage a differential degradation attack, the adversary can block a channel used by the circumvention system but not the cover application.  The adversary can also create network conditions that cause a significant drop in the application-level performance of the circumvention system but not the cover application, rendering the former unusable without collateral damage to the latter. 

\section{Disrupting Snowflake}
\label{sec:snowflake}

As a warm-up exercise, we demonstrate a simple differential degradation attack on the Snowflake pluggable transport for Tor.  The attack exploits a mismatch between how Snowflake and popular WebRTC-based applications use WebRTC channels.  

There are many other known attacks on Snowflake (see Section~\ref{sec:related}).  The purpose of this section to illustrate the principle that \emph{tunneled circumvention systems can be blocked without detection, fingerprinting, or traffic analysis}.

\subsection{Discrepancy with cover application}

The WebRTC data channel is a \emph{synchronous} peer-to-peer channel (see Section~\ref{sec:webrtc}).  Both peers must be online, i.e., running an application or keeping a browser tab open.   This is a good match for Snowflake, which transports TCP- and TLS-based Tor connections.

Unlike Snowflake, popular Web applications that use WebRTC need \emph{asynchronous} channels for their non-media data.  For example, a user of Discord or Google Meet should be able to send text messages to offline users who do not currently have the application open to receive them. Therefore, these applications cannot use the WebRTC data channel.  Instead, they typically send non-media data to central servers over TLS or QUIC and notify the receivers so receivers can retrieve the messages once they are online.

\subsection{Overview of the attack}

We present a simple attack that blocks Snowflake without protocol fingerprinting or traffic analysis. The adversary only needs to inspect a constant number of bytes in a single network packet and to keep minimal information for each connection.  This attack is thus available even to very weak real-world censors (see Section~\ref{sec:threat}), who cannot perform a complete deep inspection of network packets and do not have the capacity to compute and keep complex statistical information for every connection.

The attack consists of two steps: 1) identifying a WebRTC connection; and 2) blocking its data channel.  

\para{Identifying WebRTC connection.}
\label{sec:identify-webrtc}
WebRTC uses self-signed certificates to authenticate peers.  During signaling, a peer sends the fingerprint of his certificate to the remote peer using SDP.  During the DTLS key exchange, the latter checks if the received certificate matches the fingerprint provided in SDP.

All WebRTC self-signed certificates have the same issuer string “WebRTC” and appear in plaintext in key exchange messages.  Therefore, if a network observer sees a DTLS key exchange with a certificate that has the “WebRTC” issuer, he knows that the DTLS connection in question belongs to a WebRTC session.

\para{Blocking WebRTC data channel.}
The goal of the attack is to block the WebRTC data channel without affecting the media channel.  Recall from Section~\ref{sec:webrtc} that the media channel uses DTLS only for key exchange.  No media data are sent in DTLS record-layer “application\_data” packets.  On the other hand, all data-channel communications are sent in DTLS “application\_data” packets.  Since DTLS packet type can be identified by inspecting the packet payload, a network adversary can selectively drop DTLS “application\_data” packets to block the data channel.

\begin{figure}[h]
\centering
\includegraphics[scale=0.2]{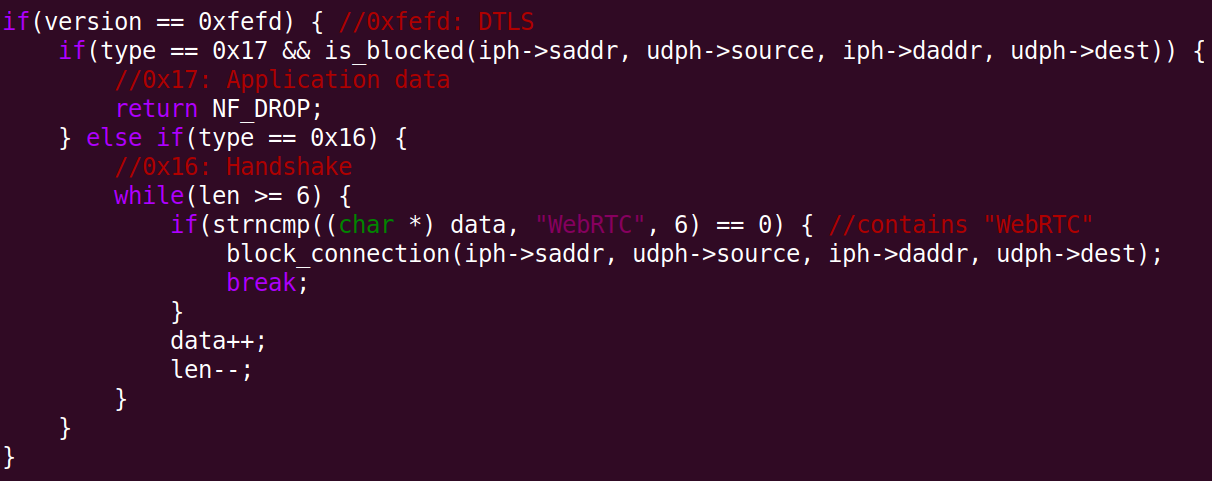}
\centering
\caption{Attack code.}
\label{fig:attack-code}
\end{figure}

\subsection{Implementing the attack}

Our proof-of-concept attack uses the Netfilter framework provided by the Linux kernel to inspect and drop network packets.  Figure~\ref{fig:attack-code} shows the code. For each DTLS key exchange message, we first check if it is a WebRTC connection by checking if the certificate issuer is "WebRTC". If the connection is a WebRTC connection, we add the [src\_ip, src\_port, dst\_ip, dst\_port] tuple to a dictionary. For each DTLS “applicaton\_data” packet, we drop the packet if the [src\_ip, src\_port, dst\_ip, dst\_port] tuple is in the dictionary.

\begin{figure}[h]
\centering
\includegraphics[width=\linewidth]{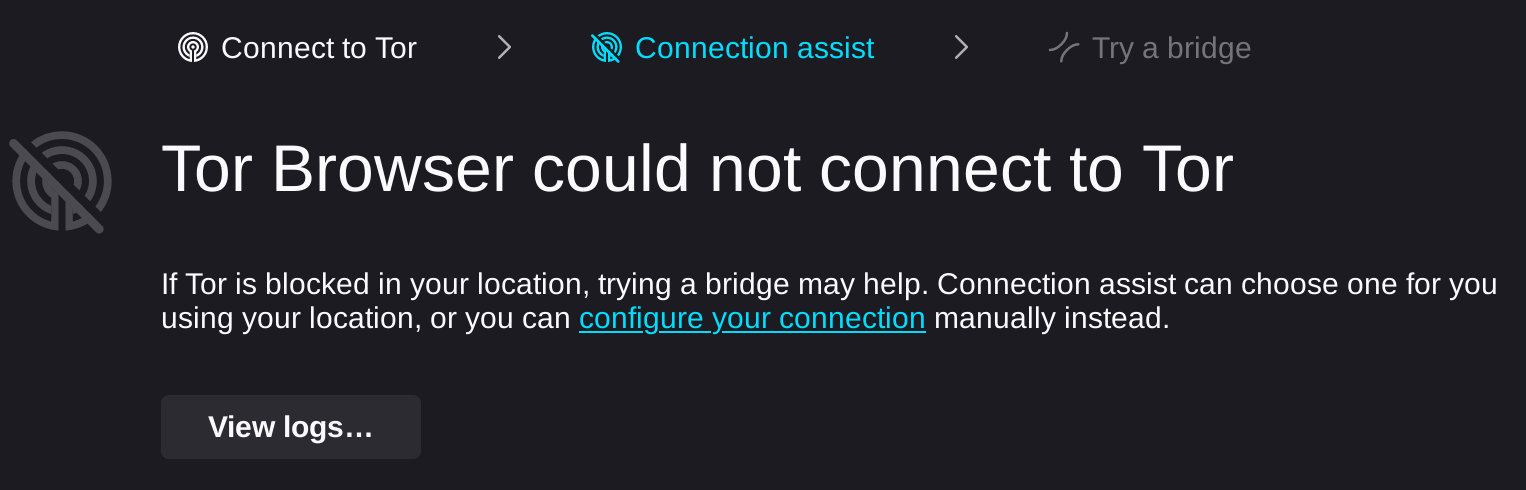}
\centering
\caption{Tor fails to connect to Snowflake.}
\label{fig:snowflake-blocked}
\end{figure}

\begin{figure}[h]
\centering
\includegraphics[width=\linewidth]{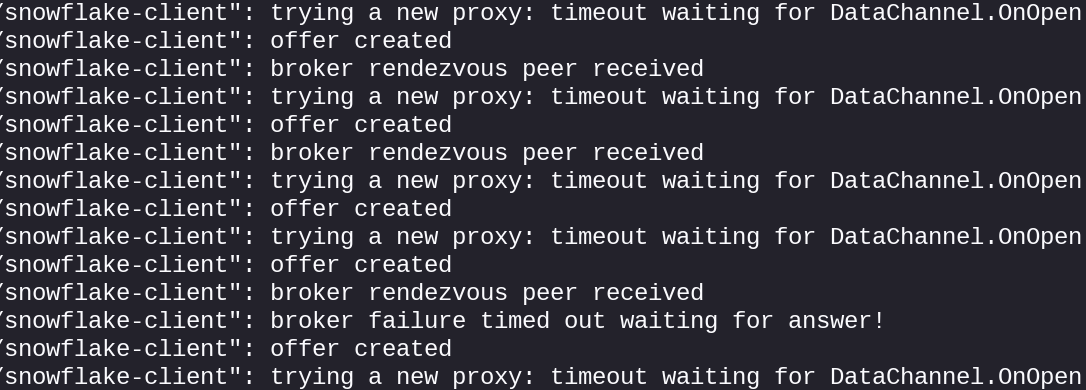}
\centering
\caption{Snowflake cannot connect to a proxy.}
\label{fig:snowflake-log}
\end{figure}

Figure \ref{fig:snowflake-blocked} demonstrates that Tor fails to connect to Snowflake under our attack. Tor logs (Figure \ref{fig:snowflake-log}) show that WebRTC is able to connect to the peer but the data channel cannot be established.

\subsection{Collateral damage}
\label{sec:snowflake-collateral}

To evaluate the collateral damage of our attack, we check if WebRTC-based applications function with the data channel blocked.  Four popular applications\textemdash Discord, Google Chat/Meet, Facebook Messenger, and Slack\textemdash use WebRTC as their video streaming protocol.  For each application, we test five functionalities: sending text messages to an online user and an offline user, sending files, audio and video calling. Table~\ref{table:snowflake-collateral} shows that none of the functionalities are affected if the data channel is blocked.

We also evaluate the video-streaming performance of Discord with and without the attack.  For a (1080p, 24fps) video, average fps (frames per second) is $23.7\pm1.4$ with the attack and $23.2\pm2.3$ without.  We conclude that blocking the WebRTC data channel does not affect the performance of video streaming.

The only WebRTC-based applications we found that do
use the data channel are file-sharing applications. Snapdrop~\cite{snapdrop} and ShareDrop~\cite{sharedrop} are browser-based applications, similar to Apple’s Airdrop.  Both send data only within the local network.  Their WebRTC data channels would not cross the network boundary where censors operate in our threat model (see Section~\ref{sec:threat}) and would not be affected by our attack. FilePizza~\cite{filepizza} is a relatively obscure application that enables users to directly share files.  When the sender initiates a transfer, the application generates a link that recipients use to download the file over the WebRTC data channel.  Since the sender and receiver may be in a free and censored regions, respectively, file transfer could be mistakenly blocked by the censor.

We conclude that ISP-level blocking of WebRTC data channels would completely block Snowflake, at the cost of potentially also blocking some file-sharing applications, but without any impact on Discord, Google Chat/Meet, Facebook Messenger or Slack.

\begin{table}[h]
\smaller
\centering
\begin{tabular}{|l|c|c|c|c|}
\hline
Applications       & \begin{tabular}[c]{@{}c@{}}Text message \\(online/offline)\end{tabular} & Audio call & Video call & File transfer \\ \hline
Discord            & \checkmark/\checkmark         & \checkmark & \checkmark & \checkmark    \\ \hline
\begin{tabular}[c]{@{}l@{}}Google \\Chat/Meet\end{tabular}   & \checkmark/\checkmark         & \checkmark & \checkmark & \checkmark    \\ \hline
\begin{tabular}[c]{@{}l@{}}Facebook \\Messenger\end{tabular} & \checkmark/\checkmark         & \checkmark & \checkmark & \checkmark    \\ \hline
Slack              & \checkmark/\checkmark         & \checkmark & \checkmark & \checkmark    \\ \hline
\end{tabular}
\caption{Snowflake attack has no impact on the functionality of WebRTC-based applications.}
\label{table:snowflake-collateral}

\begin{tabular}[c]{@{}c@{}} \end{tabular}
\end{table}

\section{Disrupting Protozoa}
\label{sec:protozoa}

Protozoa~\cite{barradas2020poking} is a behaviorally independent~\cite{raven-popets2022} system that tunnels through the media channel of a WebRTC-based video-call application by substituting video traffic at the transport layer.  

\subsection{Discrepancy with cover application}

The WebRTC media channel is unreliable. Protozoa needs reliability (to support tasks such as Web browsing), and thus runs TCP over the channel.  We demonstrate a differential degradation attack that exploits this mismatch to disrupt Protozoa\textemdash without identifying Protozoa flows and therefore without violating behavioral independence.  Furthermore, for some choices of the cover video's resolution, the attack can also cause Protozoa to break behavioral independence and become easily detectable (see Appendix \ref{sec:detectprotozoa}).

\subsection{Overview of the attack}
\label{sec:protozoa-desync}

The goal is to create network conditions under which Protozoa suffers performance degradation that makes it unusable but video quality of a genuine WebRTC video call remains acceptable. 

The naive approach, similar to the desynchronization
attack in~\cite{geddes2013cover}, is to introduce high packet losses.  It does not work against Protozoa.  In fact,
it has the \emph{opposite} effect: a 10\% packet loss
causes a tear-down of the video stream, while Protozoa preserves throughput of 160 Kbps as reported in~\cite{barradas2020poking}.

\begin{figure}[h]
\centering
\includegraphics[width=\linewidth]{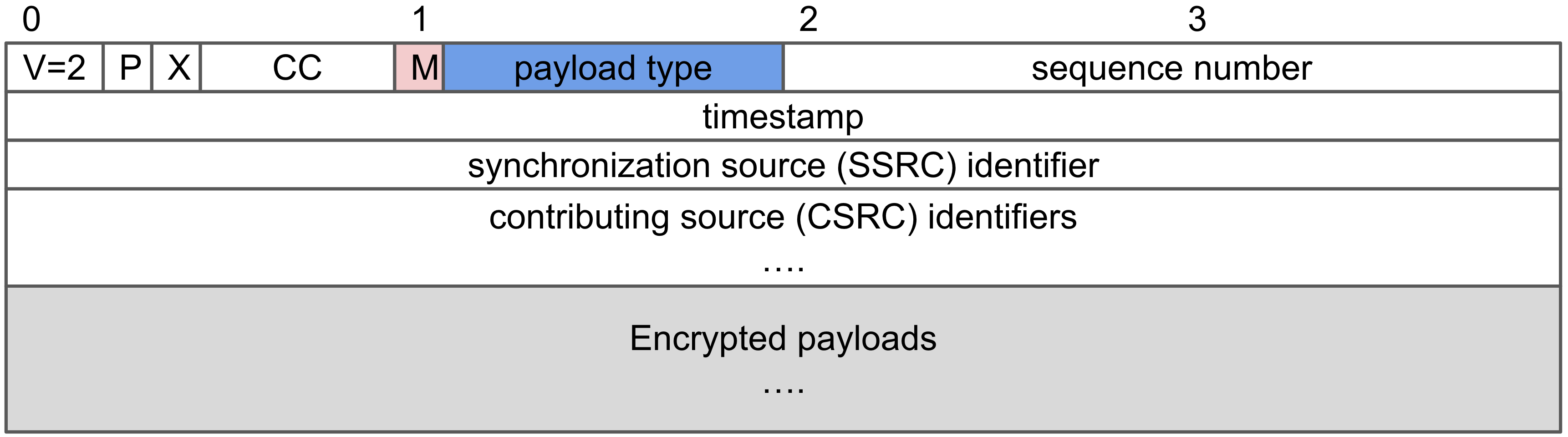}
\centering
\caption{Format of SRTP packets.  The Marker bit and the payload type field are red and blue, respectively.}
\label{fig:srtp}
\end{figure}

Our key observation is that the packet-loss attack need not apply to all packets equally.  Because a video frame usually consists of multiple packets, a single lost packet may cause the loss of an entire frame.  Therefore, the frame-loss rate of the video application is higher than the packet-loss rate.  Protozoa messages, however are usually much smaller than a video frame.  Therefore, the message-loss rate of Protozoa is lower than the frame-loss rate.  If the attacker can drop \emph{entire frames} (as opposed to single packets) without affecting other frames, he can increase Protozoa's message-loss rate while minimizing the frame-loss rate of the video application.

Figure \ref{fig:srtp} shows the format of SRTP packets used by the WebRTC media channel to transmit video content.  The payload is encrypted but the plaintext header contains enough information to identify packets that belong to a video stream, as well as video-frame boundaries within the stream.  An application-aware adversary can operate at the granularity of a video frame (rather than network packet) and selectively drop frames from the media channel. 

The attack can be staged by a constant-packet network adversary (see Section~\ref{sec:threat}).  It consists of the following steps: 

\para{Identifying WebRTC media channel.} We use the same method as Section~\ref{sec:identify-webrtc} to identify the WebRTC media channel.

\para{Identifying WebRTC video stream.} Video and audio streaming usually share the same media channel. Since Protozoa only replaces video data, it is necessary to identify video packets to avoid collateral damage to the audio stream.  The payload type field in the SRTP header reveals this information: there are only two payload types (102 and 77) that correspond to video data. 

\para{Identifying WebRTC video frames.} For video streams, the Marker bit in the SRTP header indicates video-frame boundaries (i.e., the end of the frame). Therefore, the attacker can segment a WebRTC video stream into individual frames and selectively drop any frame. 

This attack does not require long-term or multi-flow measurements. The adversary only needs to store constant-size information for each connection, to keep track if the connection is a WebRTC channel and if the next video frame should be dropped or not.  The attacker does not need to perform any measurements that involve multiple packets, nor flow classification, nor machine learning.

\subsection{Experimental setup}

\begin{figure}[h]
\centering
\includegraphics[width=\linewidth]{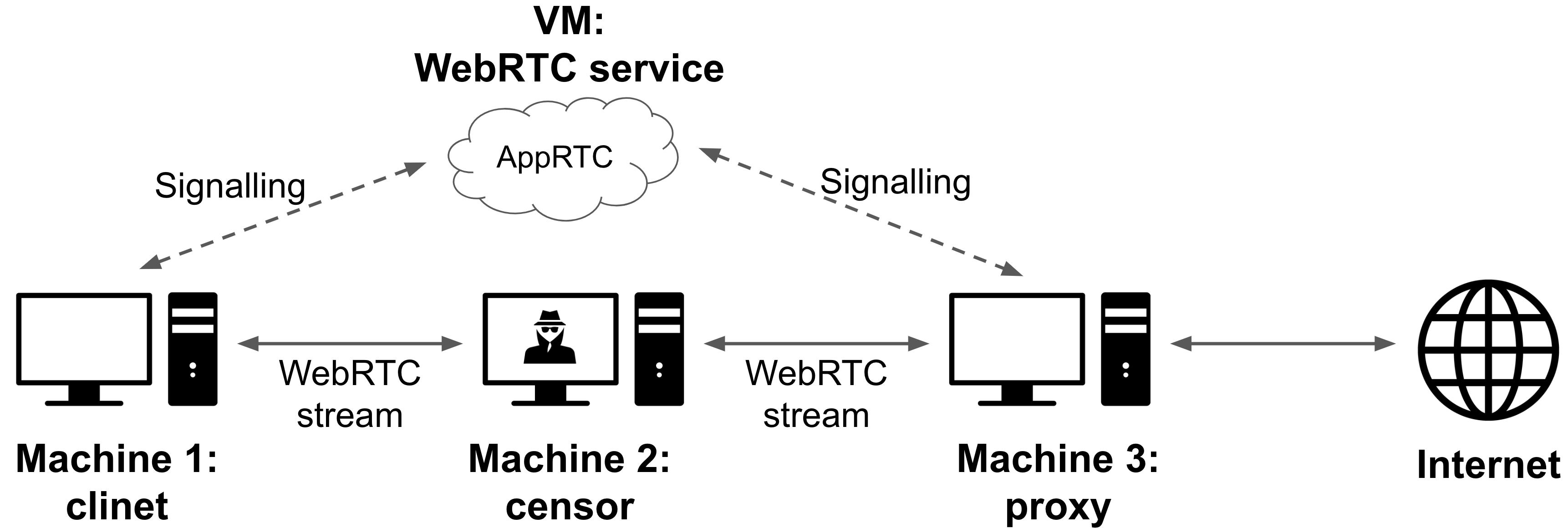}
\centering
\caption{Experimental setup.}
\label{fig:setup}
\end{figure}

Figure \ref{fig:setup} shows the setup of our experimental testbed.  We use an unmodified Protozoa implementation (commit 1c8a99f) from its GitHub repo~\cite{protozoa}.  We strictly followed the instructions in the README file to set up the environment and execute Protozoa.

Our Protozoa client is running on an Intel Xeon W-1370P @ 3.6GHz with 8 cores and 32GB of RAM (machine 1).  The Protozoa proxy is running on an Intel Core i7-7700 @ 3.6GHz with 4 cores and 16GB of RAM (machine 3).  Machines 1 and 3 are connected to a ``censor'' middlebox (machine 2) that has an Intel Core i7-5960X @ 3GHz with 8 cores and 96G of RAM.  The middlebox acts as a gateway and router for both machine 1 and machine 3. Machines 1 and 3 run 64-bit Ubuntu 20.04 LTS,  machine 2 runs 64-bit Ubuntu 16.04 LTS. Machine 3 is connected to the internet.

We use the Linux Netfilter framework~\cite{netfilter} to implement kernel modules that inspect packets, collect traces, and drop packets.  We use the Linux traffic control tool \texttt{tc} to introduce network latency.

We use AppRTC~\cite{apprtc} as the WebRTC service for our experiments.  It was evaluated in the Protozoa paper~\cite{barradas2020poking} and achieves similar performance to other services.  AppRTC is Google’s open-sourced demo application based on WebRTC.  It was served via \url{https://appr.tc} which is no longer available.  Instead, we built our own AppRTC service from the source code and ran it in a VM.  Protozoa client and proxy connect to this VM only for signaling. All WebRTC media streams still go through the censor middlebox. 

For all experiments, we use Big Buck Bunny~\cite{bigbuckbunny}, a publicly available short animation film as the input of the WebRTC video source. The original video is in MOV format with 24 fps, 1080p resolution and H.264 encoding. We also converted the video into different resolutions (720p, 540p, 360p, and 240p). All video files are played using the \texttt{ffmpeg} video library with the \texttt{v4l2loopback} kernel module to emulate a camera for the video call.

\subsection{Microbenchmarks}
\label{sec:microbenchmarks}

We measure several microbenchmarks to evaluate the performance of Protozoa under different frame-loss rates.  We use \texttt{iPerf} to measure throughput and \texttt{ping} to measure network round trip time.  For each measurement, we run the test 10 times and compute the average. Table \ref{table:micro-1080} shows how Protozoa performs with the baseline (1080p, 24fps) video.  Throughput drops to less than 7 KBytes/s when the frame-loss rate exceeds 15\%, and RTT greatly increases to around 6 seconds due to the increase of the jitter. Protozoa's message-loss rate is relatively low (< 5\%) when the frame-loss rate is less than or equal to 15\%.  We conjecture that this is due to the quality-of-service (QoS) mechanism in WebRTC such as NACK and FEC~\cite{webrtcfec}. As the frame-loss rate increases to 25\%, Protozoa's message-loss rate grows close to the frame-loss rate. 

To measure collateral damage to video-call streams, we use \textit{chrome://webrtc-internals}, a built-in dashboard in the Chrome browser that measures WebRTC statistics. Table \ref{table:micro-1080} shows video-stream performance under different frame-loss rates.  Unlike packet loss, which tears down the video stream when it exceeds 10\%, frame loss doesn't interrupt the video stream.  When the frame-loss rate is 5\%, there is no resolution drop and the reduction in the number of frames per second is less than 10\%.  The video stream still works even when the frame-loss rate is 25\%.  With higher frame-loss rates, resolution drops from $1920\times1080$ to $960\times540$, but the average number of frames per second is still over 50\% of the original video.  We deem this video performance usable.

Table~\ref{table:micro-240} shows the results for lower-resolution, (240p, 24fps) videos.   Throughput of Protozoa is lower, but it can tolerate much higher frame-loss rate and has much lower RTT.  We conjecture that the QoS mechanism of WebRTC works better when video resolution is low and jitter is smaller.  Low-resolution videos also tolerate higher frame-loss rate. The video stream maintains more than 80\% of the original video's fps with no resolution drop when the frame-loss rate is 50\%.  The stream works even when the frame-loss rate is 70\%: average fps is still more than 50\% of the original video and resolution drops slightly from $426\times240$ to $320\times180$. 

Adding extra network latency to the frame-drop attack causes a similar degradation in Protozoa's performance with less collateral damage to the video stream. Table~\ref{table:micro-240-200ms} shows the same microbenchmarks for the 240p video with the network RTT of 200ms.  When the frame-loss rate is 45\% with extra latency, performance of Protozoa is similar to 70\% loss without extra latency.  Quality of the video stream (19 fps) is higher than without extra latency (12 fps).

\begin{table}[h]
\small
\centering
\begin{tabular}{|r|ccc|cc|}
\hline
\multirow{2}{*}{\begin{tabular}[c]{@{}c@{}}Frame\\ Loss\\ Rate\end{tabular}} & \multicolumn{3}{c|}{Protozoa performance}                                                                                                                                       & \multicolumn{2}{c|}{Video stream performance}                                                                                                \\ \cline{2-6} 
                                                                             & \multicolumn{1}{c|}{\begin{tabular}[c]{@{}c@{}}Throughput\\ (KBytes/s)\end{tabular}} & \multicolumn{1}{c|}{\begin{tabular}[c]{@{}c@{}}Latency\\ RTT (ms)\end{tabular}} & Loss   & \multicolumn{1}{c|}{\begin{tabular}[c]{@{}c@{}}Average\\ FPS\end{tabular}} & \begin{tabular}[c]{@{}c@{}}Stream\\ Resolution\end{tabular} \\ \hline
$0\%$                                                                        & \multicolumn{1}{c|}{$151.0\pm14.4$}                                                  & \multicolumn{1}{c|}{$195\pm45$}                                                 & $0\%$  & \multicolumn{1}{c|}{$24.0\pm0.4$}                                          & $1920\times1080$                                            \\ \hline
$5\%$                                                                        & \multicolumn{1}{c|}{$30.8\pm9.2$}                                                    & \multicolumn{1}{c|}{$3087\pm894$}                                               & $4\%$  & \multicolumn{1}{c|}{$22.0\pm5.0$}                                            & $1920\times1080$                                            \\ \hline
$15\%$                                                                       & \multicolumn{1}{c|}{$7.0\pm5.5$}                                                     & \multicolumn{1}{c|}{$5994\pm676$}                                               & $5\%$  & \multicolumn{1}{c|}{$18.6\pm7.0$}                                            & $960\times540$                                              \\ \hline
$25\%$                                                                       & \multicolumn{1}{c|}{$4.7\pm3.7$}                                                     & \multicolumn{1}{c|}{$6002\pm967$}                                               & $28\%$ & \multicolumn{1}{c|}{$13.4\pm7.0$}                                            & $960\times540$                                              \\ \hline
\end{tabular}
\caption{Microbenchmarks.  Values are in avg$\pm$stdev format.  Source video resolution: $1920\times1080$.}
\label{table:micro-1080}
\end{table}

\begin{table}[h]
\small

\centering
\begin{tabular}{|r|ccc|cc|}
\hline
\multirow{2}{*}{\begin{tabular}[c]{@{}c@{}}Frame\\ Loss\\ Rate\end{tabular}} & \multicolumn{3}{c|}{Protozoa performance}                                                                                                                                       & \multicolumn{2}{c|}{Video stream performance}                                                                                                \\ \cline{2-6} 
                                                                             & \multicolumn{1}{c|}{\begin{tabular}[c]{@{}c@{}}Throughput\\ (KBytes/s)\end{tabular}} & \multicolumn{1}{c|}{\begin{tabular}[c]{@{}c@{}}Latency\\ RTT (ms)\end{tabular}} & Loss   & \multicolumn{1}{c|}{\begin{tabular}[c]{@{}c@{}}Average\\ FPS\end{tabular}} & \begin{tabular}[c]{@{}c@{}}Stream\\ Resolution\end{tabular} \\ \hline
$0\%$                                                                        & \multicolumn{1}{c|}{$136.3\pm10.8$}                                                  & \multicolumn{1}{c|}{$184\pm41$}                                                 & $0\%$  & \multicolumn{1}{c|}{$24.0\pm0.5$}                                          & $426\times240$                                              \\ \hline
$30\%$                                                                       & \multicolumn{1}{c|}{$10.4\pm1.6$}                                                    & \multicolumn{1}{c|}{$545\pm229$}                                                & $0\%$  & \multicolumn{1}{c|}{$24.0\pm2.4$}                                          & $426\times240$                                              \\ \hline
$50\%$                                                                       & \multicolumn{1}{c|}{$6.4\pm1.6$}                                                     & \multicolumn{1}{c|}{$1291\pm379$}                                               & $2\%$  & \multicolumn{1}{c|}{$21.5\pm5.5$}                                          & $426\times240$                                              \\ \hline
$70\%$                                                                       & \multicolumn{1}{c|}{$2.9\pm1.6$}                                                     & \multicolumn{1}{c|}{$2315\pm871$}                                               & $11\%$ & \multicolumn{1}{c|}{$12.4\pm9.3$}                                          & $320\times180$                                              \\ \hline
\end{tabular}
\caption{Microbenchmarks.  Source video resolution: $426\times240$.}
\label{table:micro-240}

\end{table}

\begin{table}[h]
\small
\centering
\begin{tabular}{|r|ccc|cc|}
\hline
\multirow{2}{*}{\begin{tabular}[c]{@{}c@{}}Frame\\ Loss\\ Rate\end{tabular}} & \multicolumn{3}{c|}{Protozoa performance}                                                                                                                                       & \multicolumn{2}{c|}{Video stream performance}                                                                                                             \\ \cline{2-6} 
                                                                             & \multicolumn{1}{c|}{\begin{tabular}[c]{@{}c@{}}Throughput\\ (KBytes/s)\end{tabular}} & \multicolumn{1}{c|}{\begin{tabular}[c]{@{}c@{}}Latency\\ RTT (ms)\end{tabular}} & Loss   & \multicolumn{1}{c|}{\begin{tabular}[c]{@{}c@{}}Average\\ FPS\end{tabular}} & \begin{tabular}[c]{@{}c@{}}Stream\\ Resolution\end{tabular}              \\ \hline
$0\%$                                                                        & \multicolumn{1}{c|}{$112.1\pm9.3$}                                                   & \multicolumn{1}{c|}{$318\pm71$}                                                 & $0\%$  & \multicolumn{1}{c|}{$24.0\pm0.6$}                                          & $426\times240$                                                           \\ \hline
$15\%$                                                                       & \multicolumn{1}{c|}{$12.3\pm1.8$}                                                    & \multicolumn{1}{c|}{$593\pm174$}                                                & $0\%$  & \multicolumn{1}{c|}{$23.4\pm4.1$}                                          & $426\times240$                                                           \\ \hline
$30\%$                                                                       & \multicolumn{1}{c|}{$5.5\pm1.7$}                                                     & \multicolumn{1}{c|}{$1501\pm478$}                                               & $4\%$  & \multicolumn{1}{c|}{$21.2\pm5.6$}                                          & \begin{tabular}[c]{@{}c@{}}$426\times240$\\ /$320\times180$\end{tabular} \\ \hline
$45\%$                                                                       & \multicolumn{1}{c|}{$3.1\pm1.43$}                                                    & \multicolumn{1}{c|}{$2189\pm711$}                                               & $22\%$ & \multicolumn{1}{c|}{$19.3\pm9.1$}                                          & $320\times180$                                                           \\ \hline
\end{tabular}
\caption{Microbenchmarks, $200$ms network RTT.  Source video resolution: $426\times240$.}
\label{table:micro-240-200ms}

\end{table}

\subsection{Disrupting Protozoa (without detection)}
\label{sec:web-browsing}

To evaluate the effect of the attack at the application level, we measure performance of Web browsing over Protozoa.  We use the baseline (1080p, 24fps) video and set the frame-loss rate to 25\% with no extra latency.  We use a Firefox browser and the Protozoa proxy to browse the front pages of five popular websites that are blocked by censors in multiple countries.  We browse each page 10 times and measure load times; timeout is set to 900 seconds for each page. 

Table \ref{table:web-1080} shows the results. Under the attack, page load time increases 6x to 23x. Some pages take more than 500 seconds to load.  More than 30\% of page loads fail (vs.\ $0$ without the attack) due to the timeout or broken connection.

As Tables \ref{table:micro-240} and \ref{table:micro-240-200ms} show, for videos with very low resolution (240p), video streaming remains operational under severe network conditions, such as 70\% frame-loss rate or 45\% frame-loss rate with 200ms RTT.  Table~\ref{table:web-240} shows the Web browsing performance of Protozoa under these conditions.  With 45\% frame-loss rate and 200ms RTT, CNN always fails to load; New York Times fails most of the time (8/10); other webpages take around 500 seconds to load.  The results are similar for 70\% frame-loss rate and no extra latency.

We conclude that Protozoa's application-level performance degrades so much that the system is effectively unusable.  Genuine video streaming remains operational.

\begin{table}[h]
\small
\centering
\begin{tabular}{|l|cc|cc|}
\hline
\multirow{2}{*}{Website} & \multicolumn{2}{c|}{No loss}                                                          & \multicolumn{2}{c|}{$25\%$ Frame Loss}                                                \\ \cline{2-5} 
                         & \multicolumn{1}{c|}{\begin{tabular}[c]{@{}c@{}}Load Time (s)\end{tabular}} & Fails  & \multicolumn{1}{c|}{\begin{tabular}[c]{@{}c@{}}Load Time (s)\end{tabular}} & Fails  \\ \hline
\href{en.wikipedia.org}{wikipedia}         & \multicolumn{1}{c|}{$10.0\pm2.5$}                                             & $0/10$ & \multicolumn{1}{c|}{$232.7\pm75.0$}                                          & $3/10$ \\ \hline
\href{bbc.com}{bbc}                  & \multicolumn{1}{c|}{$51.2\pm33.8$}                                           & $0/10$ & \multicolumn{1}{c|}{$492.1\pm338.2$}                                         & $3/10$ \\ \hline
\href{cnn.com}{cnn}                  & \multicolumn{1}{c|}{$92.7\pm18.6$}                                           & $0/10$ & \multicolumn{1}{c|}{$546.9\pm205.9$}                                         & $6/10$ \\ \hline
\href{nytimes.com}{nytimes}              & \multicolumn{1}{c|}{$86.7\pm28.1$}                                           & $0/10$ & \multicolumn{1}{c|}{$530.3\pm147.8$}                                         & $4/10$ \\ \hline
\href{reddit.com}{reddit}               & \multicolumn{1}{c|}{$56.6\pm10.3$}                                           & $0/10$ & \multicolumn{1}{c|}{$441.6\pm127.4$}                                         & $3/10$ \\ \hline
\end{tabular}
\caption{Web browsing performance of Protozoa.  Source video resolution: $1920\times1080$.}
\label{table:web-1080}
\end{table}

\begin{table}[h]
\small
\centering
\begin{tabular}{|l|cc|cc|}
\hline
\multirow{2}{*}{Website} & \multicolumn{2}{c|}{70\% Frame Loss}                                                  & \multicolumn{2}{c|}{\begin{tabular}[c]{@{}c@{}}$45\%$ Frame Loss +\\ 200ms RTT\end{tabular}} \\ \cline{2-5} 
                         & \multicolumn{1}{c|}{\begin{tabular}[c]{@{}c@{}}Load Time (s)\end{tabular}} & Fails  & \multicolumn{1}{c|}{\begin{tabular}[c]{@{}c@{}}Load Time (s)\end{tabular}}    & Fails      \\ \hline
\href{en.wikipedia.org}{wikipedia}         & \multicolumn{1}{c|}{$197.4\pm79.0$}                                          & $0/10$ & \multicolumn{1}{c|}{$264.0\pm124.6$}                                            & $1/10$     \\ \hline
\href{bbc.com}{bbc}                  & \multicolumn{1}{c|}{$494.4\pm83.6$}                                          & $2/10$ & \multicolumn{1}{c|}{$504.6\pm199.8$}                                            & $1/10$     \\ \hline
\href{cnn.com}{cnn}                  & \multicolumn{1}{c|}{$843.8\pm24.9$}                                          & $7/10$ & \multicolumn{1}{c|}{-}                                                          & $10/10$    \\ \hline
\href{nytimes.com}{nytimes}              & \multicolumn{1}{c|}{$507.1\pm253.2$}                                         & $3/10$ & \multicolumn{1}{c|}{$481.2\pm12.7$}                                             & $8/10$     \\ \hline
\href{reddit.com}{reddit}               & \multicolumn{1}{c|}{$527.2\pm151.0$}                                         & $0/10$ & \multicolumn{1}{c|}{$543.6\pm189.7$}                                            & $3/10$     \\ \hline
\end{tabular}
\caption{Web browsing performance of Protozoa.  Source video resolution: $426\times240$.}
\label{table:web-240}

\end{table}

\section{Resisting Differential Degradation Attacks}
\label{section:resisting}

When designing a tunneled circumvention system, the choice of the cover application matters.  If the circumvention system is intended to support real-time, interactive, online communications such as Web browsing, the primary requirement is that the cover application provide a high-bandwidth, low-latency, bidirectional, encrypted network channel.

\begin{figure}[t]
\centering
\includegraphics[width=\linewidth]{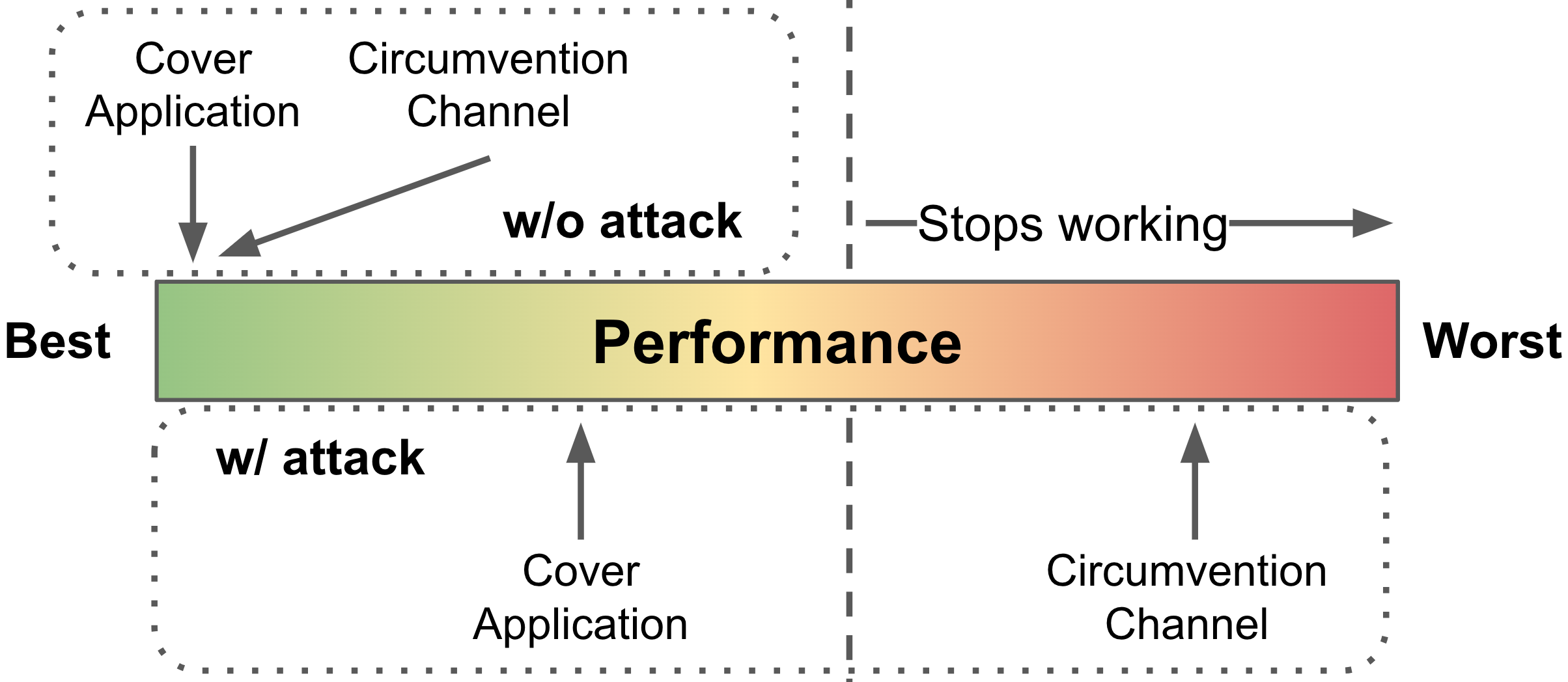}
\centering
\caption{Channel is vulnerable to differential degradation.}
\label{fig:vulnerable}
\end{figure}

The censor may not be ready to block this channel entirely, to avoid collateral damage to the users of the cover application, but he can still control the channel's quality of service. As Section~\ref{sec:protozoa} shows, the adversary can exercise this control to elicit differences in the application-level behavior of the circumvention system and the cover application, in order to damage and effectively block the former with minimal collateral damage to the latter.  Intuitively, when the adversary degrades network conditions, the circumvention system ``suffers'' before the cover application does (see Figure~\ref{fig:vulnerable}).

\begin{figure}[t]
\centering
\includegraphics[width=\linewidth]{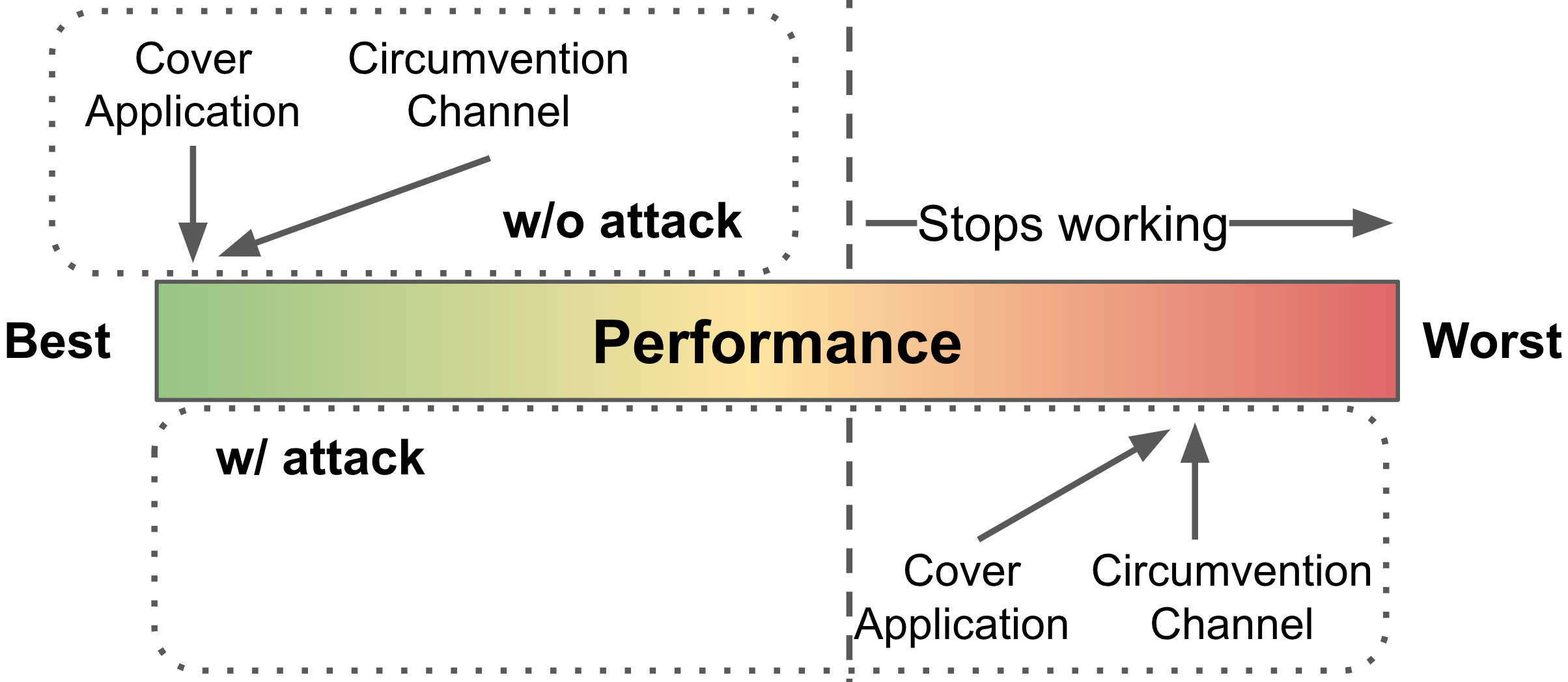}
\centering
\caption{Channel resists differential degradation.}
\label{fig:resistant}
\end{figure}

To resist differential degradation attacks, it is important that the circumvention system tolerate poor network quality of service as well as, or better than, the cover application. When the adversary degrades network conditions, the cover application should ``suffer'' before the circumvention system does (see Figure~\ref{fig:resistant}), resulting in collateral damage before circumvention is blocked.

If the circumvention system is intended to resist more advanced adversaries capable of multi-flow traffic analysis (see Section~\ref{sec:threat}), there is an additional requirement that the system \emph{not create new or unusual network flows}, in comparison to the flows typically associated with the cover application.

\subsection{Tradeoffs}
\label{sec:tradeoffs}

\begin{figure}[t]
\centering
\includegraphics[width=\linewidth]{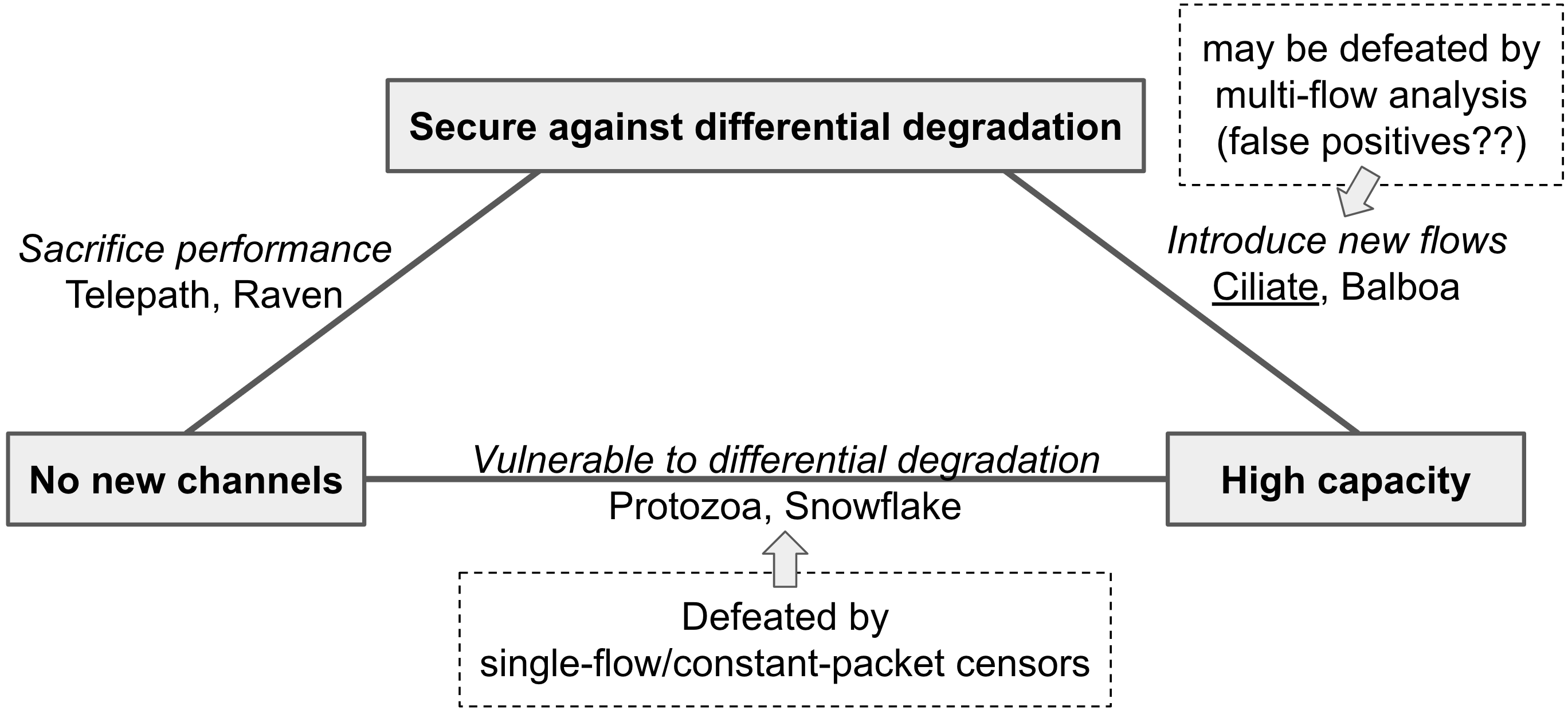}
\centering
\caption{Tradeoffs in designing circumvention systems. The three requirements cannot be satisfied at the same time with the cover applications available today.}
\label{fig:tradeoff}
\end{figure}

No cover applications available today (see Section~\ref{sec:cover}) satisfy all three requirements.  Tunneled circumvention systems must choose between vulnerability to differential degradation, introducing new network flows, or sacrificing performance (see Figure~\ref{fig:tradeoff}).

\para{Remain vulnerable to differential degradation.}
Interactive-video applications provide a high-bandwidth, low-latency, bidirectional channel but they use WebRTC, the protocol specifically designed for this functionality.  As explained in Section~\ref{sec:protozoa}, WebRTC headers reveal the application, enabling the adversary to craft application-specific differential degradation attacks.  Furthermore, quality-of-service requirements of interactive video are very different from the requirements of circumvention functionalities such as Web browsing.  By design, interactive-video applications can adjust resolution and adapt even to very poor network conditions that prevent Web browsing from working.

\para{Introduce new flows.}
The alternative to WebRTC-based cover applications is to use applications based on TLS or QUIC.  Unlike WebRTC, network traffic transmitted over these protocols does not immediately reveal which application generated it.  Identifying the application from observations of TLS or QUIC requires traffic analysis (as opposed to single-packet inspection of WebRTC headers).  This significantly raises the costs and technical difficulty for the adversary, who would need to collect large amounts of traffic data, correctly label it, and train classifiers to recognize specific applications~\cite{tong2018novel,de2018identifying}.  
Further, these classifiers are potentially susceptible to false positives, increasing the risk of collateral damage.

Within the set of candidate cover applications that use TLS or QUIC, the quality-of-service requirements of the chosen application should match those of the circumvention system.  For example, video streaming over TLS can be a suitable cover application. In Section~\ref{sec:ciliate}, we present a proof of concept that modifies Protozoa to run over video streaming.

TLS- or QUIC-based video-streaming applications provide high-capacity network channels, but these channels are \emph{unidirectional}. (Bidirectional channels are typically used in interactive applications and thus implemented using WebRTC.)  This is a mismatch for circumvention. Supporting Web browsing over video-streaming applications requires \emph{two} channels, which may create unusual network flows\textemdash in particular, the circumvention system would require an upstream, recipient-to-sender flow, which may not be common for typical uses of video streaming.  Reliably detecting anomalous flows requires a good model of ``normal'' TLS flows and multi-flow traffic analysis.  This significantly raises the bar for the adversary vs.\ differential degradation attacks (see Section~\ref{sec:threat}) and increases the risk of false positives with subsequent collateral damage to non-circumvention users of video-streaming applications.

\para{Sacrifice performance.}
Another point in the design space of circumvention systems is to pick a popular cover application that uses TLS or QUIC and provides an encrypted, bidirectional channel.  This excludes video applications, which are either unidirectional, or implemented over WebRTC.  Substituting content in non-video applications requires understanding of the application's semantics, to avoid violating application-specific invariants observable even in encrypted network traffic~\cite{sun2023telepath}.

Moving even higher in the protocol stack, a circumvention system can use steganographic techniques to encode its data in the application's content~\cite{mcpherson2016covertcast, figueira2022stegozoa}.  These systems generally have lower capacity. Systems with extra undetectability properties, such as Raven~\cite{raven-popets2022}, which provides human-level behavior indistinguishability, have even worse performance which is not sufficient for any practical circumvention task. Their resistance to differential degradation attacks is an interesting topic for future work.

\subsection{Ciliate: Protozoa over TLS}
\label{sec:ciliate}

In this section, we describe our proof-of-concept implementation of Ciliate, a variant of Protozoa that runs over TLS and thus avoids WebRTC's vulnerabily to differential degradation attacks.

\begin{figure}[t]
\centering
\includegraphics[width=\linewidth]{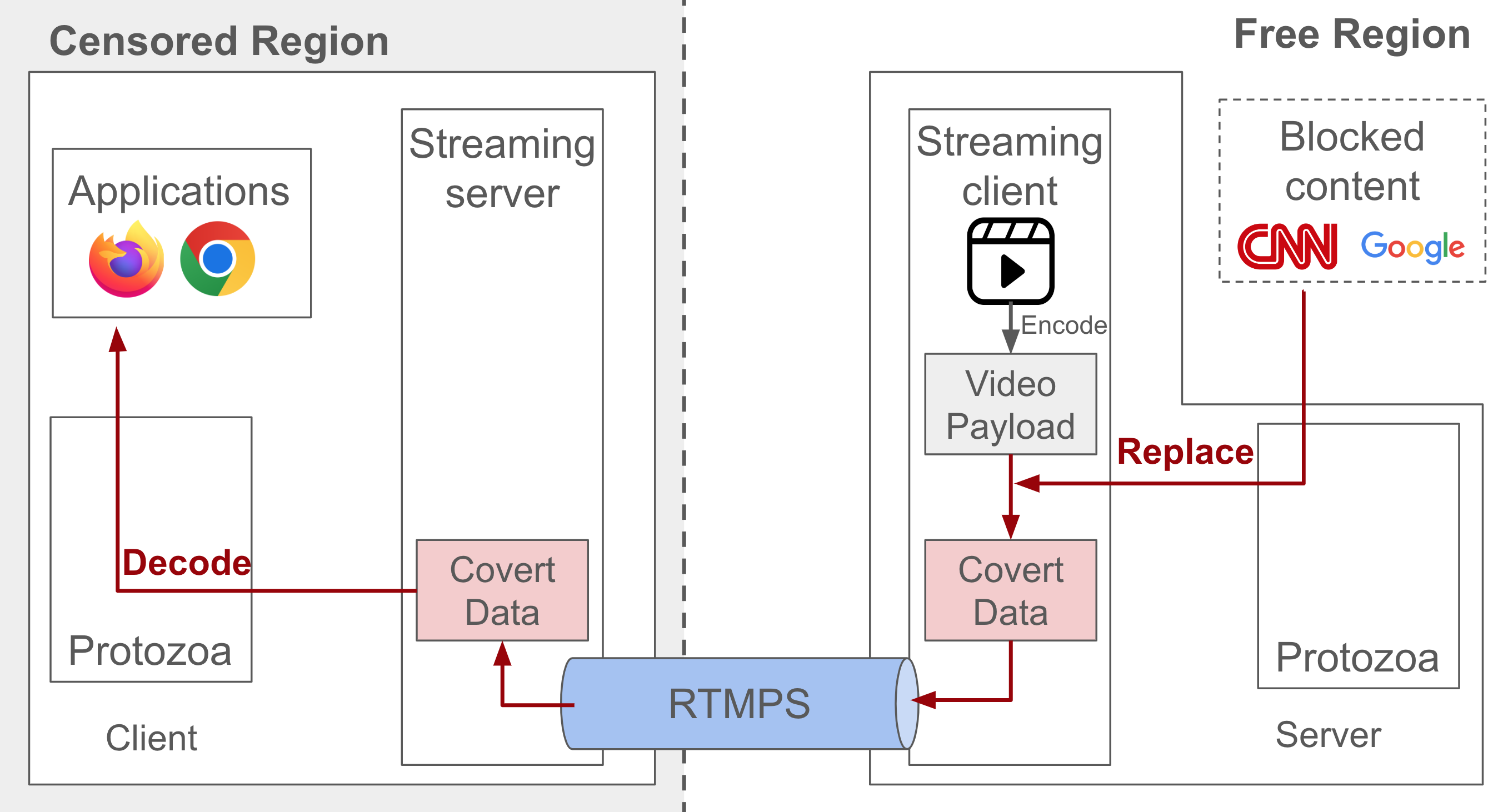}
\centering
\caption{Overview of Ciliate.}
\label{fig:ciliate}
\end{figure}

\para{Design.}
Ciliate keeps the Protozoa architecture but changes the cover application from video calls based on WebRTC to live-streaming applications based on Real-Time Messaging Protocol over TLS (RTMPS).  These applications allows users to publish a video to a server, which viewers watch via this server in real time. 

Figure \ref{fig:ciliate} shows an overview of the Ciliate design.  As in Protozoa, the streaming client first encodes the video, then replaces its content with circumvention data
and sends it to a streaming server through a RTMPS channel. The modified payload is decoded by the Protozoa code.  This design provides a unidirectional channel only: the live-streaming client is the sender, the streaming server the receiver.  To support bidirectional communication, Ciliate runs two channels as described above but in opposite directions.  This may create an unusual pattern of network flows, but detecting it requires multi-flow analysis (see Section~\ref{sec:tradeoffs}).  This is the same point in the design space (Figure~\ref{fig:tradeoff}) as chosen by Balboa~\cite{rosen2021balboa}.

Our implementation keeps most Protozoa code unmodified and replaces the WebRTC application with a modified open-source media server called MediaMTX~\cite{mediamtx}. MediaMTX allows users to read and publish RTMP(S) video streams at the same time, so it can be used both as a streaming client and a streaming server. Our modification to MediaMTX is around 350 lines of Go code. 

\para{Evaluation.}
We evaluate how our proof-of-concept implementation of Ciliate resists differential degradation attack. Since RTMPS streams data over TLS, there is no plaintext information that would enable the adversary to identify the specific application and/or recognize individual video frames. Therefore, the frame-loss attack is not available and we use the packet-loss attack for evaluation.  Results show that both Ciliate and the cover application work under 5\% packet loss; neither works under 7\% packet loss. This satisfies the collateral damage requirement discussed in section \ref{section:resisting}.

We also evaluate performance of Ciliate under 5\% packet loss to show that the system is still usable. We use the same microbenchmarks as in Section~\ref{sec:microbenchmarks} and the same Web browsing measurements as in Section~\ref{sec:web-browsing}. Table \ref{table:micro-def} shows the microbenchmarks. With 5\% packet loss, Ciliate achieves higher throughput than Protozoa \emph{without} an attack. Latency of Ciliate under the attack is higher than latency of Protozoa without an attack, but much lower than latency of Protozoa under the attack. Table \ref{table:web-def} shows Web browsing results.  Page load time of Ciliate under the attack is less than 2x of Protozoa's load time without an attack. No website fails to load with Ciliate. Therefore, we conclude that Ciliate achieves resistance to differential degradation attacks.

\begin{table}[h]
\centering
\begin{tabular}{|c|c|c|c|}
\hline
\begin{tabular}[c]{@{}c@{}}Packet Loss\end{tabular} & \begin{tabular}[c]{@{}c@{}}Throughput\\ (KBytes/s)\end{tabular} & \begin{tabular}[c]{@{}c@{}}Latency\\ RTT (ms)\end{tabular} & Loss  \\ \hline
5\%                                                   & $226.8\pm61.0$                                                  & $442.9\pm350.2$                                            & $0\%$ \\ \hline
\end{tabular}
\caption{Microbenchmarks of Ciliate.  Source video resolution: $1920\times1080$.}
\label{table:micro-def}
\end{table}


\begin{table}[h]
\centering
\begin{tabular}{|l|cc|}
\hline
\multirow{2}{*}{Website} & \multicolumn{2}{c|}{5\% packet loss}                                                   \\ \cline{2-3} 
                         & \multicolumn{1}{c|}{\begin{tabular}[c]{@{}c@{}}Load Time (s)\end{tabular}} & Fails  \\ \hline
\href{en.wikipedia.org}{wikipedia}         & \multicolumn{1}{c|}{$13.7\pm4.6$}                                           & $0/10$ \\ \hline
\href{bbc.com}{bbc}                  & \multicolumn{1}{c|}{$65.6\pm19.7$}                                           & $0/10$ \\ \hline
\href{cnn.com}{cnn}                  & \multicolumn{1}{c|}{$178.9\pm50.3$}                                          & $0/10$ \\ \hline
\href{nytimes.com}{nytimes}              & \multicolumn{1}{c|}{$161.2\pm52.6$}                                          & $0/10$ \\ \hline
\href{reddit.com}{reddit}               & \multicolumn{1}{c|}{$63.5\pm14.2$}                                           & $0/10$ \\ \hline
\end{tabular}
\caption{Web browsing performance of Ciliate under 5\% packet loss.  Source video resolution: $1920\times1080$.}
\label{table:web-def}
\end{table}

\para{Tradeoff.}
Protozoa and Ciliate illustrate the fundamental tradeoff shown in Figure~\ref{fig:tradeoff}.  Protozoa uses a WebRTC-based cover application and is thus vulnerable to detection-free differential degradation. 
Recent studies~\cite{Wu2023a} show that detection-free disruption has already been adopted by real-world censors.

Ciliate uses a TLS-based cover application and\textemdash since popular TLS-based streaming applications are unidirectional\textemdash needs to introduce new flows to support interactive functionality.  It is thus potentially vulnerable to detection.  Reliably detecting anomalous network flows without false positives requires much more sophisticated censorship capabilities than differential degradation (see Section~\ref{sec:threat}) and raises the bar for the censors vs.\ Protozoa. 

Unfortunately, with the existing cover applications that are popular today in censorship regions, no tunneled circumvention system can simultaneously (a) support high bandwidth and bidirectional communication, (b) resist differential degradation, and (c) not introduce new network flows\textemdash see Section~\ref{sec:tradeoffs}.

\section{Related Work}
\label{sec:related}

Surveys of application-based circumvention channels can be found in~\cite{iv2022security,khattak2016sok}.  We provide a summary in Section~\ref{sec:cover}.

\para{``Cover Your Acks.''}
Geddes et al.~\cite{geddes2013cover} proposed two attacks on FreeWave.  The detection attack is a multi-flow, traffic-analysis attack that trains a classifier to recognize differences in the packet-size distribution between the circumvention channel and actual speech.   The disruption attack is an early example of a differential degradation attack: it exploits the observation that FreeWave and Skype adapt differently to extremely high packet loss over a short period.  Neither attack works against Protozoa, which is behaviorally independent and adapts to high packet losses \emph{better} than its cover video-call application (see Section~\ref{sec:protozoa-desync}).

Geddes et al.\ also proposed an attack on SkypeMorph that degrades the performance of the covert channel by dropping ACK packets. Similar to~\cite{houmansadr2013parrot}, the attack exploits SkypeMorph's flawed imitation of Skype. It does not work against behaviorally independent tunneled systems (the focus of this paper) which run the actual cover application rather than imitate it.

\para{Snowflake.}
Snowflake is vulnerable to multi-flow traffic analysis~\cite{chen2023f}.  Attacks in our paper do not require any detection.

Previous work~\cite{Fifield2016b} explored fingerprinting of WebRTC-based applications by analyzing connection establishment but did not provide concrete methods to identify specific applications.  They observed that Google Hangouts and Facebook Messenger do not use WebRTC for text chat but did not explain the reason (in Section~\ref{sec:snowflake}, we show that this is due to the channel mismatch).  Finally, they mentioned that it might be possible to block Snowflake by selectively blocking the data channel and argued that tunneling data through the media channel can mitigate the attack.  This is the approach implemented by Protozoa, and we demonstrate that it can still be blocked without detection (see Section~\ref{sec:protozoa}).

In a concurrent, independent work~\cite{Bocovich2024a}, the authors of Snowflake acknowledge that other WebRTC applications do not utilize the data channel.  The paper does not analyze the prevalence of data-channel use in popular, real-world WebRTC applications (see Section~\ref{sec:snowflake-collateral}), which is essential to validate the assumption that blocking the data channel would not result in significant collateral damage.

\para{Tunneled circumvention.}
Turbo Tunnel~\cite{fifield2020turbo} is a design pattern for circumvention systems that uses an interior session and a reliability layer over some possibly unreliable obfuscation layer. Tunneling reliable protocols over unreliable channels may create mismatches that would make the system vulnerable to differential degradation.  Instantiations of Turbo Tunnel must be designed very carefully, only choosing the channels whose QoS matches the circumvention system, or else channels that can carry different applications without revealing the specific application. 

Balboa~\cite{rosen2021balboa} proposed a framework for using existing TLS-based applications as tunneled circumvention channels.  Its resistance to differential degradation depends on the implementation.  Balboa relies on a static, pre-determined traffic model to emulate application content,
thus it effectively requires cover applications to produce the same network traffic under arbitrary network conditions.  We believe that the two instantiations in~\cite{rosen2021balboa} resist differential degradation.  Neither supports bidirectional communication. 

Stegozoa~\cite{figueira2022stegozoa} encodes circumvention data into video content using steganography and streams the video through WebRTC.  Since there is no traffic substitution in Stegozoa, its application-level behavior should be the same as the cover application and thus resist differential degradation. The capacity of the system is too low to support functionality such as Web browsing.

Some recent circumvention systems include application semantics and even user behavior in their threat model, assuming they can be observed in encrypted network traffic. Telepath~\cite{sun2023telepath} avoids observable violations of application semantics by substituting only the content that is independent of the application state. Raven~\cite{raven-popets2022} uses generative adversarial networks (GANs) to imitate user behavior. Evaluation of both systems only considers machine learning-based traffic analysis.  In both systems, application- and user-level indistinguishability come at the cost of low channel capacity.

\section{Conclusion}

Over the past decade, research on censorship circumvention, as well as real-world experience, demonstrated that systems that evade IP address-based blocking can still be vulnerable to detection by network-based censors.  In response, designers of circumvention systems have been putting out increasingly elaborate proposals that attempt to hide circumvention-related activities in the network traffic of popular cover applications, such as video streaming.  These proposed systems use the same network protocols as cover applications, or even execute the latter and replace application content with circumvention data before it is transmitted.  The goal is ``behavioral independence'': all traffic sent or received by the circumvention system is indistinguishable from the traffic that would have been sent or received by a cover application under typical usage.

In this paper, we showed that even this, very ambitious indistinguishability is insufficient.  An adversary in control of the network can degrade network conditions in a way that elicits different application-level behavior from the circumvention system and the cover application.  The circumvention system then suffers a catastrophic drop in performance resulting in effective blocking\textemdash even if this degradation is not observable by the adversary.  Critically, the cover application does not suffer a corresponding drop in performance, thus a censor can deploy a differential degradation attack without the risk of collateral damage to non-circumvention users of the cover application.

The root cause of differential degradation vulnerabilities is the mismatch between the circumvention systems' and cover applications' respective requirements for their network channels.  Using Snowflake and Protozoa as concrete examples, we showed how a network-based adversary can exploit (a) application-level information revealed by WebRTC, and (b) mismatches in channel usage or adaptation to changing network conditions between these systems and WebRTC-based video applications they use for cover.

We explained the tradeoffs faced by designers of circumvention systems given the limited choice of cover applications, and proposed Ciliate, a modification of Protozoa that uses TLS-based video-streaming for cover and resists differential degradation.

The main lesson of this paper is: \emph{cover application matters!}  It is difficult and perhaps impossible to design a circumvention system in a way that is agnostic to the semantics and network requirements of the cover application, since the discrepancy \emph{at any level}\textemdash even application-level adaptations and performance changes that are not observable from the network\textemdash can make a system vulnerable to detection and/or blocking.  Therefore, designers of circumvention systems should choose cover applications that match the intended functionality of their systems and understand how the chosen application would behave in a wide range of conditions.

\para{Acknowledgements.} Supported by DARPA and AFRL under
Contract FA8750-19-C-0079.

\bibliographystyle{abbrv}
\bibliography{main}

\appendix
\section{Detecting Protozoa}
\label{sec:detectprotozoa}

The main purpose of this paper is to demonstrate that  simple, single-flow, differential degradation attacks can disrupt and block tunneled circumvention systems without needing to distinguish circumvention and cover traffic.  Detection is not a primary goal of differential degradation.  In our experiments with Protozoa, however, we noticed that our frame-loss attack causes Protozoa to generate a distinguishable traffic pattern when using high-resolution WebRTC videos for cover.  For completeness, we report these observations in this appendix.

Microbenchmark results in Section \ref{sec:microbenchmarks} show that, when the frame-loss is high, WebRTC adaptively reduces resolution of the streamed video.   Frames of a lower-resolution video are smaller than those of a higher-resolution video.

\begin{figure}[h]
\centering
\includegraphics[width=\linewidth]{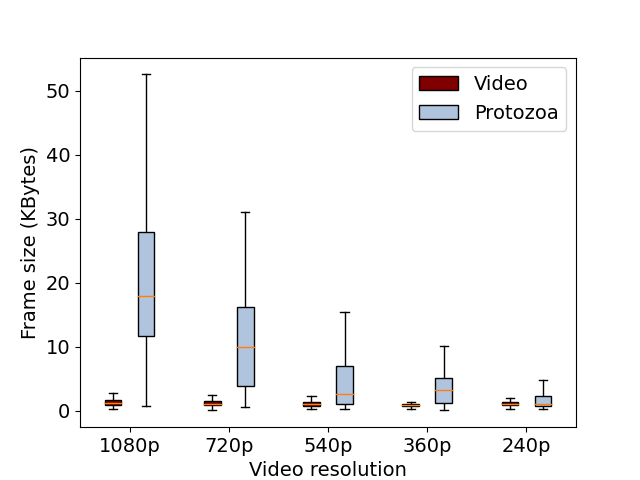}
\centering
\caption{Frame-size distribution with videos of different resolutions under 25\% frame loss.}
\label{fig:frame-size}
\end{figure}

Nevertheless, the implementation of Protozoa available from the GitHub repo does not change the resolution of its cover video under any network conditions.   As a result, under the frame-loss attack, Protozoa's video frames are much larger than the frames of a genuine WebRTC video stream.  Figure~\ref{fig:frame-size} shows, for all video resolutions, the respective frame-size distributions of Protozoa and a WebRTC video stream when the frame-loss rate is 25\%.  With any resolution other than 240p, Protozoa frames can be easily distinguished from genuine video-stream frames. A simple, single-feature, binary distinguisher\textemdash any flow that does not reduce frame sizes under the frame-loss attack is not a genuine WebRTC flow\textemdash can identify Protozoa flows with perfect accuracy.  This distinguisher does not require any multi-flow measurements or machine learning.

The root cause of this behavior is not clear.  WebRTC congestion control is based on the information from frame headers.  According to the design in~\cite{barradas2020poking}, Protozoa does not replace frame headers and thus, in theory, should be able to adjust the sending resolution in response to network changes.  Empirically, the publicly available implementation of Protozoa does not adjust, creating an easily recognizable traffic pattern. 

\end{document}